\documentclass[twocolumn,twocolappendix]{aastex631}

\usepackage{showyourwork}
\usepackage{amsmath}
\usepackage{amssymb}
\usepackage{graphicx}
\usepackage{xcolor}

\usepackage{siunitx}
\usepackage{xspace}

\newcommand{\yes}{\textcolor{green}{\checkmark}\xspace}
\newcommand{\meh}{\textcolor{black}{$\sim$}\xspace}
\newcommand{\no}{\textcolor{red}{$\times$}\xspace}
\newcommand{\osuaffil}{Department of Astronomy, The Ohio State University, 140 W. 18th Ave, Columbus OH 43210, USA}
\newcommand{\ccappaffil}{Center for Cosmology and AstroParticle Physics, The Ohio State University, 191 W. Woodruff Ave., Columbus OH 43210, USA}
\newcommand{\aFe}{[$\alpha$/Fe]\xspace}
\newcommand{\vice}{{\tt VICE}\xspace}
\newcommand{\hydro}{{\tt h277}\xspace}

\defcitealias{Johnson2021-Migration}{J21}
\defcitealias{Leung2023-Ages}{L23}

\shorttitle{SN Ia DTD in GCE Models}
\shortauthors{Dubay, Johnson, \& Johnson}

\received{11 April 2024}
\revised{8 July 2024}
\accepted{9 July 2024}
\submitjournal{ApJ}

\begin{document}

\title{Galactic Chemical Evolution Models Favor an Extended Type Ia Supernova Delay-Time Distribution}

\author[0000-0003-3781-0747]{Liam O.\ Dubay}
\affiliation{\osuaffil}
\affiliation{\ccappaffil}
\author[0000-0001-7258-1834]{Jennifer A.\ Johnson}
\affiliation{\osuaffil}
\affiliation{\ccappaffil}
\author[0000-0002-6534-8783]{James W.\ Johnson}
\affiliation{Observatories of the Carnegie Institution for Science, 813 Santa Barbara St., Pasadena CA 91101, USA}
\affiliation{\osuaffil}
\affiliation{\ccappaffil}

\correspondingauthor{Liam O.\ Dubay}
\email{dubay.11@osu.edu}

\begin{abstract}
    Type Ia supernovae (SNe Ia) produce most of the Fe-peak elements in the Universe and therefore are a crucial ingredient in galactic chemical evolution models. SNe Ia do not explode immediately after star formation, and the delay-time distribution (DTD) has not been definitively determined by supernova surveys or theoretical models. Because the DTD also affects the relationship among age, [Fe/H], and [$\alpha$/Fe] in chemical evolution models, comparison with observations of stars in the Milky Way is an important consistency check for any proposed DTD. We implement several popular forms of the DTD in combination with multiple star formation histories for the Milky Way in multi-zone chemical evolution models which include radial stellar migration. We compare our predicted interstellar medium abundance tracks, stellar abundance distributions, and stellar age distributions to the final data release of the Apache Point Observatory Galactic Evolution Experiment (APOGEE). We find that the DTD has the largest effect on the [$\alpha$/Fe] distribution: a DTD with more prompt SNe Ia produces a stellar abundance distribution that is skewed toward a lower [$\alpha$/Fe] ratio. While the DTD alone cannot explain the observed bimodality in the [$\alpha$/Fe] distribution, in combination with an appropriate star formation history it affects the goodness of fit between the predicted and observed high-$\alpha$ sequence. Our model results favor an extended DTD with fewer prompt SNe Ia than the fiducial $t^{-1}$ power law.
\end{abstract}

\section{Introduction}

Galactic chemical evolution (GCE) studies seek to explain the observed distribution of metals throughout the Milky Way Galaxy. \citet{Tinsley1979-StellarLifetimes} made a compelling case that the non-solar \aFe\footnote{
    In standard bracket notation, $[X/Y]\equiv \log_{10}(X/Y) - \log_{10}(X/Y)_{\odot}$. In this paper we will use \aFe and [O/Fe] interchangeably, although observational studies will often use a combination of $\alpha$-elements to calculate \aFe.
} ratios seen by, e.g., \citet{Wallerstein1962-GDwarfAbundances} were caused by different stellar lifetimes for the contributors of the Fe-peak elements than the $\alpha$-elements. Type Ia supernovae (SNe Ia), the thermonuclear explosions of carbon-oxygen white dwarfs (WDs), are responsible for a majority of the Fe produced in the Galaxy \citep{Matteucci1986-SupernovaEnrichment}; meanwhile, core collapse supernovae (CCSNe), the explosions of massive stars, produce the $\alpha$-elements (e.g., O and Mg) in addition to a smaller fraction of Fe. SNe Ia are delayed by $\sim 0.04-10$ Gyr after star formation events, as evidenced by observations in both star-forming and elliptical galaxies \citep[e.g.,][]{Maza1976-SNStatistics}. This delayed enrichment leads to a decrease in \aFe with increasing [Fe/H] \citep{Matteucci1986-SupernovaEnrichment}. Therefore, the relative abundances of the $\alpha$-elements and Fe as a function of stellar age trace the balance of SN rates over time.

The {\it delay-time distribution} (DTD) refers to the rate of SN Ia events per unit mass of star formation as a function of stellar population age \citep[for a review, see Section 3.5 of][]{Maoz2014-Review}. When the DTD is convolved with the Galactic star formation rate (SFR), it yields the overall SN Ia rate. The quantitative details of the relationship between \aFe and [Fe/H] are set by the DTD, and as such it is a key parameter in GCE models. However, the DTD remains poorly constrained because it reflects the detailed evolution of the SN Ia progenitor systems, so different models for the progenitors of SNe Ia will naturally predict different forms for the DTD.

The explosion mechanism(s) of SNe Ia are not fully understood \citep[for reviews, see][]{Maoz2014-Review,Livio2018-ProgenitorReview,Ruiter2020-ProgenitorReview,Liu2023-SNIaBinaryReview}. Two general production channels have been proposed. In the single-degenerate (SD) case, the WD accretes mass from a close non-degenerate companion until it surpasses $\sim1.4$ M$_\odot$ and explodes \citep{Whelan1973-SDModel,Nomoto1982-SDModel,Yoon2003-SDModel}. In the double-degenerate (DD) case, two WDs merge after a gravitational-wave inspiral \citep{Iben1984-IaBinary,Webbink1984-DDModel,Pakmor2012-WDMerger} or head-on collision \citep{Benz1989-CollisionalDD,Thompson2011-CollisionalDD}. Searches for signs of interaction between the SN ejecta and a non-degenerate companion \citep[e.g.,][]{Panagia2006-RadioEmission,Chomiuk2016-RadioEmission,Fausnaugh2019-EarlyIaLightCurves,Tucker2020-SNeIaSpectra,Dubay2022-SNeIaCSM} or for a surviving companion \citep[e.g.,][]{Schaefer2012-ExCompanionSNR,Do2021-Progenitor1972E,Tucker2023-SN2011fe} have placed tight constraints on the SD channel, heavily disfavoring it as the main pathway for producing ``normal'' SNe Ia. The DD channel is now the preferred model, but it faces issues with matching observed SN Ia rates because not all WD mergers necessarily lead to a thermonuclear explosion \citep[e.g.,][]{NomotoIben1985-DDMergers,SaioNomoto1998-DDMergers,Shen2012-DDMergers}, and the progenitor systems are difficult to detect even within our own Galaxy \citep{RebassaMansergas2019-WhereAreDDProgenitors}.

As a result of the uncertainty regarding SN Ia progenitors, theoretical models have yet to converge on a single prediction for the DTD. For the DD channel, assumptions about the distribution of WD separations and the rate of gravitational wave inspiral suggest a broad $\sim t^{-1}$ DTD at long delay times ($\gtrsim 1$ Gyr), but at short delays ($\lesssim 1$ Gyr) the rate is limited by the need to produce two WDs \citep[see][]{Greggio2005-AnalyticalRates,Maoz2014-Review}. Triple or higher-order progenitor systems could also produce a $t^{-1}$ DTD \citep{Fang2018-QuadrupleSystems,Rajamuthukumar2023-TripleEvolution}. The DTD which would result from the SD channel depends greatly on the assumptions of binary population synthesis, but in general is expected to cover a narrower range of delay times and may feature a steep exponential cutoff at the long end \citep[e.g.,][]{Greggio2005-AnalyticalRates}.

Surveys of SNe Ia can constrain the DTD by comparing the observed rate of SNe Ia to their host galaxy parameters \citep[e.g.,][]{Mannucci2005-SNRate,Heringer2019-FieldGalaxyDTD} or inferred star formation histories \citep[SFHs; e.g.,][]{Maoz2012-SloanIIDTD}, measuring SN Ia rates in galaxy clusters \citep[e.g.,][]{Maoz2010-ClusterDTD}, or comparing the volumetric SN Ia rate to the cosmic SFH as a whole \citep[e.g.,][]{Graur2014-VolumetricSNIaRates,Strolger2020-ExponentialDTD}. Early studies, which had limited sample sizes, produced unimodal \citep{Strolger2004-SNIaProgenitors} or bimodal \citep{Mannucci2006-TwoPopulations} DTDs where the majority of SNe Ia explode within a relatively narrow range of delay times. More recent studies have recovered broader DTD functions, with many converging on a declining power-law of $\sim t^{-1}$ \citep[e.g.,][]{Graur2013-IaRateVsMass,Graur2015-UnifiedExplanation,Maoz2017-CosmicDTD,Castrillo2021-DTD,Wiseman2021-DESRates}, though there is some evidence for a steeper slope in galaxy clusters \citep{Maoz2017-CosmicDTD,FriedmannMaoz2018-ClusterDTD}. It is especially difficult to constrain the DTD for short delay times \citep{MaozMannucci2012-SNeIaReview,Rodney2014-PromptSNeIa} because of the need for SN Ia rates at long lookback times and uncertainties in the age estimates of stellar populations.

The uncertainties in the SN Ia DTD propagate into GCE models. In principle, the observed chemical abundance patterns should therefore contain information about the DTD, and by extension the progenitors of SNe Ia. The metallicity distribution function (MDF)\footnote{In this paper, we refer to the MDF and the distribution of [Fe/H] interchangeably.} and distribution of [O/Fe] record the history of SN Ia enrichment as a function of stellar age and location in the Galaxy. A striking feature of the \aFe distribution in the Milky Way disk is the distinct separation into two components, the high- and low-$\alpha$ sequences, at similar metallicity \citep[e.g.,][]{Bensby2014-solarNeighborhoodAbundances}. Since the \aFe abundance reflects the ratio of CCSN to SN Ia enrichment, the DTD should influence the \aFe bimodality.

A few studies have investigated different DTDs in one-zone chemical evolution models, but comparisons to abundance data have been limited to the solar neighborhood \citep[e.g.,][]{Andrews2017-ChemicalEvolution,Palicio2023-AnalyticDTD}. \citet{Matteucci2009-DTDModels} compared five DTDs in a multi-zone GCE model and found that the agreement with observations worsens if the fraction of prompt ($t\lesssim 100$ Myr) SNe Ia is either too high or too low, but they were similarly limited by the available data for the solar neighborhood. \citet{Poulhazan2018-PrecisionPollution} found that the prompt component of the DTD affects the peak and width of the \aFe distribution in a cosmological smoothed-particle hydrodynamics simulation, but their simulation was not designed to reproduce the parameters of the Milky Way. The current era of large spectroscopic surveys such as the Apache Point Observatory Galactic Evolution Experiment \citep[APOGEE;][]{Majewski2017-APOGEE} and the ongoing Milky Way Mapper \citep{Kollmeier2017-SDSS-V} has made abundances across the Milky Way disk available for comparison to more sophisticated GCE models.

This paper presents a comprehensive look at the DTD in a multi-zone GCE model that can qualitatively reproduce the observed abundance structure of the Milky Way disk. A multi-zone approach allows for a radially-dependent parameterization of the SFH, outflows, stellar migration, and abundance gradient which can better match observations across the Galactic disk. We evaluate a selection of DTDs from the literature with multiple SFHs and a prescription for radial stellar migration in the Versatile Integrator for Chemical Evolution \citep[\vice;][]{JohnsonWeinberg2020-Starbursts}. In Section \ref{sec:methods}, we present our models for the DTD and SFH and describe our observational sample. In Section \ref{sec:onezone-results}, we detail our one-zone chemical evolution models and present results. In Section \ref{sec:multizone-results}, we present the results of our multi-zone models and compare to observations. In Section \ref{sec:discussion}, we discuss the implications for the DTD and future surveys. In Section \ref{sec:conclusions}, we summarize our conclusions.

\section{Methods}
\label{sec:methods}

We use \vice to run chemical evolution models which closely follow those of \citet{JohnsonWeinberg2020-Starbursts} and \citet[][hereafter \citetalias{Johnson2021-Migration}]{Johnson2021-Migration}. We refer the interested reader to the former for details about the \vice package and to the latter for details about the model Milky Way disk, including the star formation law, radial density gradient, and outflows. Similar to \citetalias{Johnson2021-Migration}, we adopt a prescription for radial migration based on the \hydro hydrodynamical simulation \citep{Christensen2012-h277}. In Appendix \ref{app:migration}, we describe our method for determining the migration distance $\Delta R_{\rm gal}$ and midplane distance $|z|$ for each model stellar population. Our method produces smoother distributions in chemical abundance space than the simulation-based approach, but the abundance distributions are otherwise unaffected by this change. Table \ref{tab:multizone-parameters} summarizes our model parameters and the sub-sections in which we discuss them in detail.

\begin{deluxetable*}{Cccl}
    \tablecaption{A summary of parameters and their fiducial values for our chemical evolution models (see discussion in Section \ref{sec:methods}). We omit some parameters that are unchanged from \citetalias{Johnson2021-Migration}; see their Table 1 for details.\label{tab:multizone-parameters}}
    \tablehead{
        \colhead{Quantity} & \colhead{Fiducial Value(s)} & \colhead{Section} & \colhead{Description}
    }
    \startdata
        R_{\rm gal}     & [0, 20] kpc   & \ref{sec:multizone-results} & Galactocentric radius \\
        \delta R_{\rm gal}  & 100 pc    & \ref{sec:multizone-results} & Width of each concentric ring \\
        \Delta R_{\rm gal}  & N/A       & \ref{app:migration} & Change in orbital radius due to stellar migration \\
        p(\Delta R_{\rm gal}|\tau,R_{\rm form}) & Equation \ref{eq:radial-migration}    & \ref{app:migration} & Probability density function of radial migration distance \\
        z                   & [-3, 3] kpc                & \ref{app:migration} & Distance from Galactic midplane at present day \\
        p(z|\tau,R_{\rm final}) & Equation \ref{eq:sech-pdf}            & \ref{app:migration} & Probability density function of Galactic midplane distance\\
        \Delta t        & 10 Myr    & \ref{sec:multizone-results} & Time-step size \\
        t_{\rm max}     & 13.2 Gyr  & \ref{sec:multizone-results} & Disk lifetime \\
        n               & 8         & \ref{sec:multizone-results} & Number of stellar populations formed per ring per time-step \\
        R_{\rm SF}      & 15.5 kpc  & \ref{sec:multizone-results} & Maximum radius of star formation \\
        M_{g,0}   & 0         & \ref{sec:sfh}     & Initial gas mass \\
        \dot M_r    & continuous    & \ref{sec:multizone-results} & Recycling rate \citep[][Equation 2]{JohnsonWeinberg2020-Starbursts} \\
        \hline
        R_{\rm Ia}(t)   & Equation \ref{eq:dtd-function}    & \ref{sec:dtd-models}  & delay-time distribution of Type Ia supernovae \\
        t_D             & 40 Myr    & \ref{sec:dtd-models}  & Minimum SN Ia delay time \\
        N_{\rm Ia}/M_\star  & $2.2\times10^{-3}$ M$_\odot^{-1}$ & \ref{sec:yields}  & SNe Ia per unit mass of stars formed \citep{MaozMannucci2012-SNeIaReview} \\
        \hline
        y_{\rm O}^{\rm CC}  & 0.015     & \ref{sec:yields}  & CCSN yield of O    \\
        y_{\rm Fe}^{\rm CC} & 0.0012    & \ref{sec:yields}  & CCSN yield of Fe   \\
        y_{\rm O}^{\rm Ia}  & 0         & \ref{sec:yields}  & SN Ia yield of O       \\
        y_{\rm Fe}^{\rm Ia} & 0.00214   & \ref{sec:yields}  & SN Ia yield of Fe \\
        \hline
        f_{\rm IO}(t|R_{\rm gal})   & Equation \ref{eq:insideout-sfh}   & \ref{sec:sfh} & Time-dependence of the inside-out SFR \\
        f_{\rm LB}(t|R_{\rm gal})   & Equation \ref{eq:lateburst-sfh}   & \ref{sec:sfh} & Time-dependence of the late-burst SFR \\
        \tau_{\rm rise}             & 2 Gyr     & \ref{sec:sfh} & SFR rise timescale for inside-out and early-burst models \\
        \tau_{\rm EB}(t)          & Equation \ref{eq:earlyburst-taustar}  & \ref{sec:sfh}   & Time-dependence of the early-burst SFE timescale \\
        f_{\rm EB}(t|R_{\rm gal})   & Equation \ref{eq:earlyburst-ifr}  & \ref{sec:sfh} & Time-dependence of the early-burst infall rate \\
        f_{\rm TI}(t|R_{\rm gal})   & Equation \ref{eq:twoinfall-ifr}   & \ref{sec:sfh} & Time-dependence of the two-infall infall rate \\
        \hline
        \tau_\star                    & 2 Gyr & \ref{sec:onezone-results} & SFE timescale in one-zone models \\
        \eta(R_{\rm gal}=8\,{\rm kpc})  & 2.15  & \ref{sec:onezone-results} & Outflow mass-loading factor at the solar annulus \\
        \tau_{\rm sfh}(R_{\rm gal}=8\,{\rm kpc})    & 15.1 Gyr  & \ref{sec:sfh} & SFH timescale at the solar annulus \\
    \enddata
\end{deluxetable*}
\vspace{-24pt}

\subsection{Nucleosynthetic Yields}
\label{sec:yields}

For simplicity and easier comparison to the results of \citetalias{Johnson2021-Migration}, we focus our analysis on O and Fe, representing the $\alpha$ and Fe-peak elements, respectively. Both elements are produced by CCSNe. \vice adopts the instantaneous recycling approximation for CCSNe, so the equation which governs CCSN enrichment as a function of star formation for some element $x$ is simply
\begin{equation}
    \dot M_x^{\rm CC}(t) = y_x^{\rm CC} \dot M_\star(t)
    \label{eq:ccsn-enrichment}
\end{equation}
where $y_x^{\rm CC}$ is the CCSN yield of element $x$ per unit mass of star formation, and $\dot M_\star$ is the SFR. Following \citetalias{Johnson2021-Migration}, who in turn adopt their CCSN yields from \citet{ChieffiLimongi2004-CCSNYields} and \citet{LimongiChieffi2006-CCSNYields}, we adopt $y_{\rm O}^{\rm CC}=0.015$ and $y_{\rm Fe}^{\rm CC}=0.0012$. The primary effect of these yields is to set the low-[Fe/H] ``plateau'' in [O/Fe] which represents pure CCSN enrichment. The chosen yields for this paper produce a plateau at ${\rm [O/Fe]}=0.45$; see \citet{Weinberg2023-CCSNYield} for more discussion on the effect of the CCSN yields on chemical evolution.

Following the formalism of \citet{Weinberg2017-ChemicalEquilibrium}, the rate of Fe contribution to the ISM from SNe Ia is 
\begin{equation}
    \dot M_{\rm Fe}^{\rm Ia}(t) = y_{\rm Fe}^{\rm Ia} \langle \dot M_\star\rangle_{\rm Ia}(t)\\
\end{equation}
where $\langle \dot M_\star\rangle_{\rm Ia}(t)$ is the time-averaged SFR weighted by the DTD at time $t$ and $y_{\rm Fe}^{\rm Ia}$ is the Fe yield of SNe Ia. \citet{Weinberg2017-ChemicalEquilibrium} show in their Appendix A that
\begin{equation}
    \langle \dot M_\star\rangle_{\rm Ia} \equiv \frac{\int_0^t \dot M_\star(t')R_{\rm Ia}(t-t')dt'}{\int_{t_D}^{t_{\rm max}} R_{\rm Ia}(t')dt'}
    \label{eq:weighted-sfr}
\end{equation}
where $R_{\rm Ia}$ is the DTD in units of ${\rm M}_\odot^{-1}\,{\rm yr}^{-1}$, $t_D$ is the minimum SN Ia delay time, and $t_{\rm max}$ is the lifetime of the disk. The denominator of Equation \ref{eq:weighted-sfr} is therefore equal to $N_{\rm Ia}/M_\star$, the total number of SNe Ia per M$_{\odot}$ of stars formed.

The yield $y_{\rm Fe}^{\rm Ia}$ measures the mass of Fe produced by SNe Ia over the full duration of the DTD, which can be expressed as:
\begin{equation}
    y_{\rm Fe}^{\rm Ia} = m_{\rm Fe}^{\rm Ia} \int_{t_D}^{t_{\rm max}} R_{\rm Ia}(t')dt' = m_{\rm Fe}^{\rm Ia} \frac{N_{\rm Ia}}{M_\star},
    \label{eq:dtd-integral}
\end{equation}
where $m_{\rm Fe}^{\rm Ia}$ is the average mass of Fe produced by a single SN Ia, and $N_{\rm Ia}/M_\star=2.2\pm1\times10^{-3}\,{\rm M}_\odot^{-1}$ is the average number of SNe Ia per mass of stars formed \citep{MaozMannucci2012-SNeIaReview}. Adjusting the value of $y_{\rm Fe}^{\rm Ia}$ primarily affects the end point of chemical evolution tracks in [O/Fe]--[Fe/H] space. Following \citetalias{Johnson2021-Migration}, we adopt $y_{\rm Fe}^{\rm Ia}=0.00214$. This yield is originally adapted from the W70 model of \citet{Iwamoto1999-SNIaYields}, but it is increased slightly so that the inside-out SFH produces stars with ${\rm [O/Fe]}\approx 0.0$ by the end of the model. The overall scale of the yields is inconsequential: a lower value of $y_{\rm Fe}^{\rm Ia}$ would produce similar results if compensated with a lower outflow mass-loading factor $\eta\equiv \dot M_{\rm out}/\dot M_\star$ (this is the yield-outflow degeneracy; see \citealt{Weinberg2023-CCSNYield}, \citealt{Sandford2024-StrongOutflows}, and appendix B of \citealt{Johnson2023-DwarfGalaxyArchaeology}). \citet{Palla2021-SNIaYield} studied the effect of different SN Ia yields on GCE models in detail.

\subsection{Delay-Time Distributions}
\label{sec:dtd-models}

\begin{figure}
    \centering
    \includegraphics{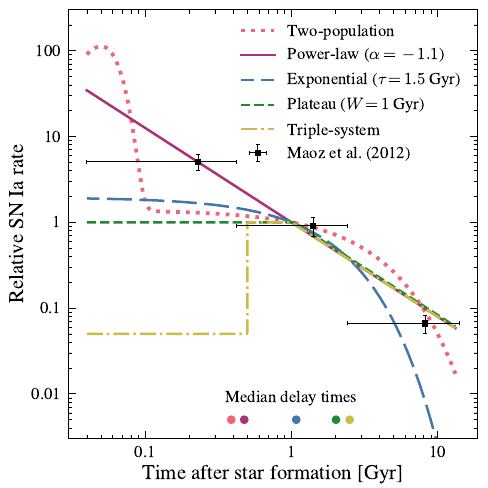}
    \caption{Selection of models for the SN Ia DTD used in this paper. All functions are normalized such that $f_{\rm Ia}(t=1\,{\rm Gyr})=1$. The black squares represent the DTD recovered for the SDSS-II sample of SNe Ia by \citet{Maoz2012-SloanIIDTD} at the same scale as the model DTDs. The horizontal and vertical error bars indicate the time range and 1$\sigma$ uncertainties of each DTD measurement, respectively. The colored circles along the horizontal axis indicate the median delay time for each model.}
    \label{fig:dtds}
    \script{delay_time_distributions.py}
\end{figure}

We explore five different functional forms for the DTD: a two-population model, a single power-law, an exponential, a broken power-law with an initially flat plateau, and a model computed from triple-system dynamics. We also investigate one or two useful variations of the input parameters for each functional form. Figure \ref{fig:dtds} presents a selection of these DTDs, and Table \ref{tab:dtds} summarizes the parameters and median delay times ($t_{\rm med}$) for all of our DTDs. We use simple forms rather than simulated physical or analytic models of SNe Ia for the sake of decreased computational time and easier interpretation of the model predictions. Physically-motivated models of the DTD must contend with many unknown or poorly-constrained parameters, so our simplified forms have the advantage of reducing the number of free parameters. In Appendix \ref{app:analytic-dtds}, we show that a few of our simple forms adequately approximate the more complete analytic models of \citet{Greggio2005-AnalyticalRates}. 

In this subsection, we present functional forms of each DTD in terms of a function $f_{\rm Ia}$ that has units of ${\rm Gyr}^{-1}$ and defines the shape of the DTD for $t\ge t_D$ as
\begin{equation}
    R_{\rm Ia}(t) = 
    \begin{cases}
        \frac{N_{\rm Ia}}{M_\star}
        \frac{f_{\rm Ia}(t)}{\int_{t_D}^{t_{\rm max}} f_{\rm Ia}(t') dt'}, & t \ge t_D \\
        0 & t < t_D.
    \end{cases}
    \label{eq:dtd-function}
\end{equation}
Note that the denominator in Equation \ref{eq:dtd-function} normalizes the DTD.

\begin{table*}
    \centering
    \caption{Summary of SN Ia DTDs explored in this paper (see discussion in Section \ref{sec:dtd-models}).}
    \label{tab:dtds}
    \begin{tabular}{lclcl}
        \hline\hline
        \multicolumn{1}{c}{Model} & Eq. & \multicolumn{1}{c}{Parameters} & $t_{\rm med}$ [Gyr] & \multicolumn{1}{c}{Similar to} \\
        \hline
        Two-population  & \ref{eq:prompt-dtd}   & $t_{\rm max}=0.05$ Gyr, $\sigma=0.015$ Gyr,   & 0.39  & \citet{Mannucci2006-TwoPopulations} \\
                        &                       & $\tau=3$ Gyr & & \\
        Power-law   & \ref{eq:powerlaw-dtd} & $\alpha=-1.4$                 & 0.18  & \citet[][cluster]{Maoz2017-CosmicDTD}; 
                                                      \citet{Heringer2019-FieldGalaxyDTD}       \\
        Power-law   & \ref{eq:powerlaw-dtd} & $\alpha=-1.1$                 & 0.48  & \citet[][field]{Maoz2017-CosmicDTD}; 
                                                      \citet{Wiseman2021-DESRates}              \\
        Exponential & \ref{eq:exponential-dtd}  & $\tau=1.5$ Gyr    & 1.08  & \citet[][SD]{Greggio2005-AnalyticalRates};
                                                                      \citet{Schonrich2009-RadialMixing};       \\
                    &                           &                   &       & \citet{Weinberg2017-ChemicalEquilibrium}  \\
        Exponential & \ref{eq:exponential-dtd}  & $\tau=3$ Gyr   & 2.08 & --- \\
        Plateau     & \ref{eq:plateau-dtd}  & $W=0.3$ Gyr, $\alpha=-1.1$    & 1.08  &\citet[][CLOSE DD]{Greggio2005-AnalyticalRates} \\
        Plateau     & \ref{eq:plateau-dtd}  & $W=1$ Gyr, $\alpha=-1.1$      & 2.02  & \citet[][WIDE DD]{Greggio2005-AnalyticalRates} \\
        Triple-system   & \ref{eq:triple-dtd}   & $f_{\rm init}=0.05f_{\rm peak}$, $t_{\rm rise}=0.5$ Gyr, & 2.50   & \citet{Rajamuthukumar2023-TripleEvolution} \\
                        &                       & $W=0.5$ Gyr, $\alpha=-1.1$ & &  \\
        \hline
    \end{tabular}
\end{table*}

\paragraph{Two-population} A DTD in which $\sim50\%$ of SNe Ia belong to a ``prompt'' Gaussian component at small $t$ and the remainder form an exponential tail at large $t$:
\begin{equation}
    f_{\rm Ia}^{\rm twopop}(t) = \frac{1}{\sigma\sqrt{2\pi}} e^{-\frac{(t-t_p)^2}{2\sigma^2}} + \frac{1}{\tau} e^{-t/\tau}.
    \label{eq:prompt-dtd}
\end{equation}
To approximate the DTD from \citet{Mannucci2006-TwoPopulations}, we take $t_p=50$ Myr, $\sigma=15$ Myr, and $\tau=3$ Gyr, which results in $\sim 40\%$ of SNe Ia exploding within $t<100$ Myr. As we illustrate in Figure \ref{fig:dtds}, the two-population DTD has a shorter median delay time than most other models (except the power-law with $\alpha=-1.4$, not shown). This formulation is slightly different than the approximation used in other GCE studies \citep[e.g.,][]{Matteucci2006-BimodalDTDConsequences,Poulhazan2018-PrecisionPollution}, where it has a more distinctly bimodal shape. We have compared the two approximations to this DTD in a one-zone model and found that they produce very similar abundance distributions. This DTD was adopted by the Feedback In Realistic Environments \citep[FIRE;][]{Hopkins2014-FIRE-1} and FIRE-2 \citep{Hopkins2018-FIRE-2} simulations.

\paragraph{Power-law} A single power law with slope $\alpha$:
\begin{equation}
    f_{\rm Ia}^{\rm plaw}(t) = (t/1\,\rm{Gyr})^\alpha
    \label{eq:powerlaw-dtd}
\end{equation}
A declining power-law with $\alpha\sim-1$ \citep{Totani2008-DTD} arises from typical assumptions about the distribution of post-common envelope separations and the rate of gravitational wave inspiral \citep[see Section 3.5 from][]{Maoz2014-Review}. It is therefore a commonly assumed DTD in GCE studies (e.g., \citealt{Rybizki2017-Chempy}; \citetalias{Johnson2021-Migration}; \citealt{Weinberg2023-CCSNYield}). Additionally, the observational evidence for a power-law DTD is strong. \citet{Maoz2017-CosmicDTD} obtained a DTD with $\alpha=-1.07\pm0.09$ based on volumetric rates and an assumed cosmic SFH for field galaxies in redshift range $0\leq z\leq 2.25$. \citet{Wiseman2021-DESRates} obtained a similar slope of $\alpha=-1.13\pm0.05$ for field galaxies in the redshift range $0.2<z<0.6$. \citet{Heringer2019-FieldGalaxyDTD} used a SFH-independent method to constrain the DTD for field galaxies within $0.01<z<0.2$ and found a larger value of $\alpha=-1.34^{+0.19}_{-0.17}$. For galaxy clusters, \citet{Maoz2017-CosmicDTD} found a steeper DTD slope of $\alpha=-1.39^{+0.32}_{-0.05}$, and \citet{FriedmannMaoz2018-ClusterDTD} found a similar slope of $\alpha=-1.3^{+0.23}_{-0.16}$; however, a re-analysis by \citet{FreundlichMaoz2021-ClusterDTD} revealed no significant difference between the cluster and field galaxy DTDs. In this paper, we investigate the cases $\alpha=-1.1$ and $\alpha=-1.4$.

\paragraph{Exponential} An exponentially declining DTD with timescale $\tau$:
\begin{equation}
    f_{\rm Ia}^{\rm exp}(t) = e^{-t/\tau}.
    \label{eq:exponential-dtd}
\end{equation}
This model allows analytic solutions to the abundances as a function of time for some SFHs, making it a popular choice \citep[e.g.,][]{Weinberg2017-ChemicalEquilibrium,Pantoni2019-AnalyticSolutions,Palicio2023-AnalyticDTD}. \citet{Schonrich2009-RadialMixing} and \citet{Weinberg2017-ChemicalEquilibrium} both assumed an exponential DTD with a timescale $\tau=1.5$ Gyr. However, observational support for an exponential DTD is scarce. \citet{Strolger2020-ExponentialDTD}, fitting to the cosmic SFH and SFHs from field galaxies, found a range of exponential-like solutions with timescales $\sim 1.5 - 6$ Gyr.
In this paper, we investigate timescales $\tau=1.5$ and 3 Gyr. We show in Appendix \ref{app:analytic-dtds} that an exponential DTD with $\tau=1.5$ Gyr is an adequate approximation for the analytic SD DTD from \citet{Greggio2005-AnalyticalRates}.

\paragraph{Plateau} A modification of the power-law in which the DTD ``plateaus'' for a duration $W$ before declining:
\begin{equation}
    f_{\rm Ia}^{\rm plat}(t) =
    \begin{cases}
        1, & t < W \\
        (t/W)^\alpha, & t \ge W.
    \end{cases}
    \label{eq:plateau-dtd}
\end{equation}
Our primary motivation is to consider a model which matches observations at delay times beyond a few Gyr, where the DTD is best constrained, but with a smaller fraction of prompt ($\lesssim 100$ Myr) SNe Ia than the single power law. This form has been used for the DTD of neutron star mergers \citep{Simonetti2019-NeutronStarDTD}, but to our knowledge it has not been considered for SNe Ia in previous GCE models. We show in Appendix \ref{app:analytic-dtds} that this form can approximate the more complicated analytic DD DTDs from \citet{Greggio2005-AnalyticalRates}. We investigate the cases $W=0.3$ Gyr and $W=1$ Gyr, taking $\alpha=-1.1$ for all plateau models.

\paragraph{Triple-system} A DTD based on simulations of triple-system evolution by \citet{Rajamuthukumar2023-TripleEvolution}. We approximate their numerically-generated DTD as a special case of the plateau model (Equation \ref{eq:plateau-dtd}) where the initial rate is quite low until an instantaneous rise to the plateau value at time $t_{\rm rise}$:
\begin{equation}
    f_{\rm Ia}^{\rm triple}(t) =
    \begin{cases}
        \epsilon, & t < t_{\rm rise} \\
        1, & t_{\rm rise} \leq t < W \\
        (t / W) ^ \alpha, & t \geq W,
    \end{cases}
    \label{eq:triple-dtd}
\end{equation}
with $t_{\rm rise}=0.5$ Gyr, $W=0.5$ Gyr, $\alpha=-1.1$, and $\epsilon=0.05$ (i.e., the initial rate is 5\% of the peak rate). As illustrated in Figure \ref{fig:dtds}, the triple-system DTD has the longest median delay time out of all the models we investigate.

There are many models for the DTD which have been proposed in the literature, and an exhaustive test of every one is infeasible. In particular, we do not consider a Gaussian DTD \citep[e.g.,][]{Strolger2004-SNIaProgenitors}, in which most SNe Ia explode with delay times close to $\sim 3$ Gyr and there are very few prompt events. Evidence for this form came from SN Ia rate measurements for $z>1$, but re-analysis of the data revealed large errors due to the small sample size, extinction corrections, and uncertainties in the SFHs \citep[e.g.,][]{Forster2006-SNIaConstraints,Greggio2008-CosmicMix}. Additionally, \citet{Mannucci2006-TwoPopulations} found that such a DTD fails to reproduce the observed dependence of the SN Ia rate on galaxy color, and there is evidence that at least some SNe Ia must be prompt in order to explain observed rates in spiral galaxies \citep[e.g.,][]{Mannucci2005-SNRate,ScannapiecoBildsten2005-SNIaRate}. GCE studies have found that a Gaussian DTD over-produces high-[O/Fe] and low-[Fe/H] stars \citep{Matteucci2009-DTDModels,Palicio2023-AnalyticDTD}.

\subsubsection{The Minimum SN Ia Delay Time}
\label{sec:minimum-delay}

In addition to the DTD shape, the minimum SN Ia delay time $t_D$ is another parameter that can have an effect on chemical evolution observables, such as the location of the high-$\alpha$ knee and the [O/Fe] distribution function \citep[DF;][]{Andrews2017-ChemicalEvolution}. The value of $t_D$ is set by the lifetime of the most massive SN Ia progenitor system. Previous GCE studies have adopted values ranging from $t_D\approx30$ Myr \citep[e.g.,][]{Poulhazan2018-PrecisionPollution} to $t_D=150$ Myr \citepalias[e.g.,][]{Johnson2021-Migration}. We take $t_D=40$ Myr as our fiducial value as it is the approximate lifetime of an $8\,{\rm M}_\odot$ star. In Section \ref{sec:onezone-results} we find that adopting a longer $t_D$ has only a minor effect on the chemical evolution for most DTDs except the power-law, but in that case the effect of a longer $t_D$ can be approximated by adding an initial plateau of width $W=0.3$ Gyr to the DTD (see Figure \ref{fig:onezone-twopanel}).

\subsection{Star Formation Histories}
\label{sec:sfh}

We consider four models for the SFH, which we refer to as inside-out, late-burst, early-burst, and two-infall. 
The former two models, which feature a smooth SFH, were investigated by \citetalias{Johnson2021-Migration} using a similar methodology to this paper. The inside-out model produced a good agreement to the age--[O/Fe] relation observed by \citet{Feuillet2019-MilkyWayAges}, while the late-burst model better matched their observed age--metallicity relation. The latter two models feature discontinuous or ``bursty'' SFHs. The early-burst model, proposed by \citet{Conroy2022-ThickDisk}, uses an efficiency-driven starburst to explain the break in the \aFe trend observed in the H3 survey \citep{Conroy2019-H3Survey}. The two-infall model was proposed by \citet{Chiappini1997-TwoInfall} and features two distinct episodes of gas infall which produce the thick and thin disks.
Together, these four models cover a range of behavior, including a smooth SFH, and SFR-, SFE-, and infall-driven starbursts.

The inside-out and late-burst models are run in \vice's ``star formation mode,'' where the SFR surface density $\dot\Sigma_\star$ is prescribed along with the star formation efficiency (SFE) timescale $\tau_\star\equiv \Sigma_g/\dot\Sigma_\star$. The remaining quantities, infall rate surface density $\dot\Sigma_{\rm in}$ and gas surface density $\Sigma_g$, are calculated from the specified quantities assuming the star formation law adopted by \citetalias{Johnson2021-Migration} (see their Equation 14). More specifically, at each timestep the infall rate is calculated to fulfill the quantity of gas required to produce the specified SFR by the star formation law. The latter two models are run in ``infall mode,'' where we specify $\dot\Sigma_{\rm in}$ and $\tau_\star$. The initial gas mass is zero for all models (including those run in star formation mode). The mode in which \vice models are run makes no difference as a unique solution can always be obtained if two of the four parametric forms are specified. 

\begin{figure*}
    \centering
    \includegraphics[width=\linewidth]{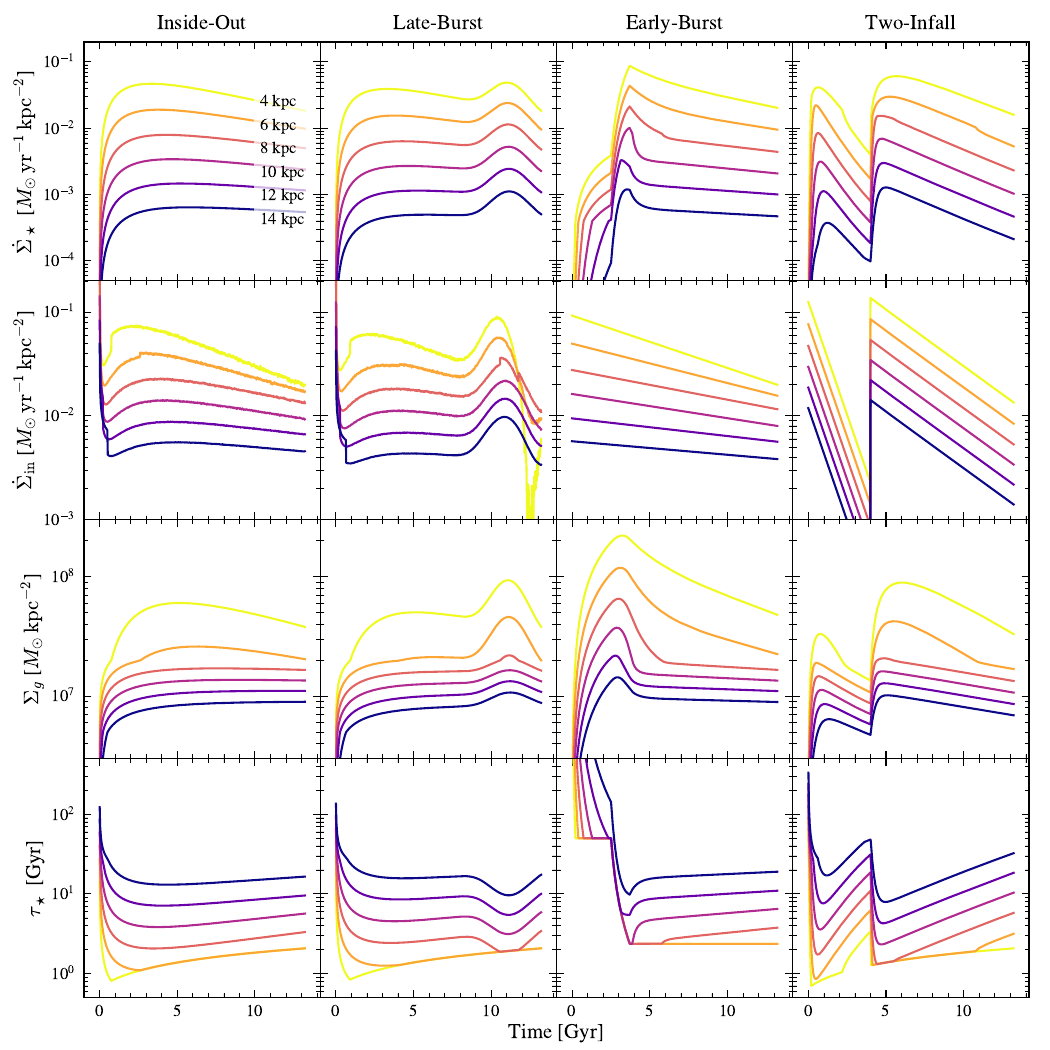}
    \caption{The surface densities of star formation $\dot \Sigma_\star$ (first row from the top), gas infall $\dot \Sigma_{\rm in}$ (second row), and gas mass $\Sigma_g$ (third row), and the SFE timescale $\tau_\star$ (fourth row) as functions of time for our four model SFHs (see discussion in Section \ref{sec:sfh}): inside-out (first column from the left; see Equation \ref{eq:insideout-sfh}), late-burst (second column; see Equation \ref{eq:lateburst-sfh}), early-burst (third column; see Equations \ref{eq:earlyburst-taustar} and \ref{eq:earlyburst-ifr}), and two-infall (fourth column; see Equation \ref{eq:twoinfall-ifr}). In each panel, we plot curves for the model zones which have inner radii at 4 kpc (yellow), 6 kpc (orange), 8 kpc (red), 10 kpc (violet), 12 kpc (indigo), and 14 kpc (blue).}
    \label{fig:sfhs}
    \script{star_formation_histories.py}
\end{figure*}

The SFH is normalized such that the model predicts a total stellar mass of $(5.17\pm1.11)\times10^{10} M_\odot$ \citep{LicquiaNewman2015-StellarMass} and the stellar surface density gradient reported by \citet[][see Appendix B of \citetalias{Johnson2021-Migration}]{BlandHawthornGerhard2016-MilkyWayReview}.
We present an overview of the four SFHs in Figure \ref{fig:sfhs}, and we discuss them individually here.

\paragraph{Inside-out} As in \citetalias{Johnson2021-Migration}, this is our fiducial SFH. The dimensionless time-dependence of the SFR is given by
\begin{equation}
    f_{\rm IO}(t|R_{\rm gal}) = \Big[1 - \exp\Big(\frac{-t}{\tau_{\rm rise}}\Big)\Big] \exp\Big(\frac{-t}{\tau_{\rm sfh}(R_{\rm gal})}\Big),
    \label{eq:insideout-sfh}
\end{equation}
where we assume $\tau_{\rm rise}=2$ Gyr for all radii. The SFH timescale $\tau_{\rm sfh}$ varies with $R_{\rm gal}$, with $\tau_{\rm sfh}(R_{\rm gal}=8\,\rm{kpc})\approx15$ Gyr at the solar annulus and longer timescales in the outer Galaxy. The $\tau_{\rm sfh} - R_{\rm gal}$ relation is based on the radial gradients in stellar age in Milky Way-like spirals measured by \citet[][see Section 2.5 of \citetalias{Johnson2021-Migration} for details]{Sanchez2020-StarFormationTimescales}.

\paragraph{Late-burst} A variation on the inside-out SFH with a burst in the SFR at late times which is described by a Gaussian according to
\begin{equation}
    f_{\rm LB}(t|R_{\rm gal}) = f_{\rm IO}(t|R_{\rm gal}) \Big(1 + A_b e^{-(t-t_b)^2/2\sigma_b^2} \Big),
    \label{eq:lateburst-sfh}
\end{equation}
where $A_b$ is the dimensionless amplitude of the starburst, $t_b$ is the time of the peak of the burst, and $\sigma_b$ is the width of the Gaussian. 
Evidence for a recent star formation burst $\sim 2-3$ Gyr ago has been found in {\it Gaia} \citep{Mor2019-Starburst} and in massive WDs in the solar neighborhood \citep{Isern2019-Starburst}.
Following \citetalias{Johnson2021-Migration}, we adopt $A_b=1.5$, $t_b=11.2$ Gyr, and $\sigma_b=1$ Gyr. The values of $\tau_{\rm rise}$ and $\tau_{\rm sfh}(R_{\rm gal})$ are the same as in the inside-out case.

\paragraph{Early-burst} An extension of the model proposed by \citet{Conroy2022-ThickDisk} to explain the non-monotonic behavior of the high-$\alpha$ sequence down to ${\rm [Fe/H]}\approx-2.5$. This model features an abrupt factor $\sim20$ rise in the SFE at early times, driving an increase in the [O/Fe] abundance at the transition between the epochs of halo and thick disk formation. \citet{Stahlholdt2022-StarFormationEpochs} found evidence for a burst $\sim10$ Gyr ago which marks the beginning of a second phase of star formation. \citet{Mackereth2018-AlphaBimodality} found that an early infall-driven burst of star formation can lead to a MW-like $\alpha$-bimodality in the EAGLE simulations \citep{Crain2015-EAGLE,Schaye2015-EAGLE}. We adopt the following formula for the time-dependence of the SFE timescale from \citet{Conroy2022-ThickDisk}:
\begin{equation}
    \frac{\tau_{\rm EB}}{1\,\rm{Gyr}} =
    \begin{cases}
        50, & t < 2.5\,\rm{Gyr} \\
        \frac{50}{[1+3(t-2.5)]^2}, & 2.5\leq t \leq 3.7\,\rm{Gyr} \\
        2.36, & t > 3.7\,\rm{Gyr}.
    \end{cases}
    \label{eq:earlyburst-taustar}
\end{equation}
While \citet{Conroy2022-ThickDisk} used a constant infall rate in their one-zone model, we adopt a radially-dependent infall rate which declines exponentially with time:
\begin{equation}
    f_{\rm EB}(t|R_{\rm gal}) = \exp\Big(\frac{-t}{\tau_{\rm sfh}(R_{\rm gal})}\Big),
    \label{eq:earlyburst-ifr}
\end{equation}
where $\tau_{\rm sfh}$ is the same as in the inside-out case.
To calculate $\dot \Sigma_\star$ from the above quantities, we modify the fiducial star formation law adopted from \citetalias{Johnson2021-Migration}, substituting $\tau_{\rm EB}$ for the SFE timescale of molecular gas:
\begin{equation}
    \dot \Sigma_\star = 
    \begin{cases}
        \Sigma_g \tau_{\rm EB}^{-1}, & \Sigma_g \geq \Sigma_{g,2} \\
        \Sigma_g \tau_{\rm EB}^{-1} \Big(\frac{\Sigma_g}{\Sigma_{g,2}}\Big)^{2.6}, & \Sigma_{g,1} \leq \Sigma_g \leq \Sigma_{g,2} \\
        \Sigma_g \tau_{\rm EB}^{-1} \Big(\frac{\Sigma_{g,1}}{\Sigma_{g,2}}\Big)^{2.6} \Big(\frac{\Sigma_g}{\Sigma_{g,1}}\Big)^{0.7}, & \Sigma_g \leq \Sigma_{g,1},
    \end{cases}
    \label{eq:earlyburst-sfr}
\end{equation}
with $\Sigma_{g,1}=5\times 10^6\,{\rm M}_\odot\,{\rm kpc}^{-2}$ and $\Sigma_{g,2}=2\times 10^7\,{\rm M}_\odot\,{\rm kpc}^{-2}$.

\paragraph{Two-infall} First proposed by \citet{Chiappini1997-TwoInfall}, this model parameterizes the infall rate as two successive, exponentially declining bursts to explain the origin of the high- and low-$\alpha$ disk populations:
\begin{equation}
    \label{eq:twoinfall-ifr}
    f_{\rm TI}(t|R_{\rm gal}) = N_1(R_{\rm gal}) e^{-t/\tau_1} + N_2(R_{\rm gal}) e^{-(t-t_{\rm on})/\tau_2},
\end{equation}
where $\tau_1=1$ Gyr and $\tau_2=4$ Gyr are the exponential timescales of the first and second infall, respectively, and $t_{\rm on}=4$ Gyr is the onset time of the second infall \citep[based on typical values in, e.g.,][]{Chiappini1997-TwoInfall,Spitoni2020-TwoInfall,Spitoni2021-TwoInfall}. $N_1$ and $N_2$ are the normalizations of the first and second infall, respectively, and their ratio $N_2/N_1$ is calculated so that the thick-to-thin-disk surface density ratio $f_\Sigma(R)=\Sigma_2(R)/\Sigma_1(R)$ is given by
\begin{equation}
    f_\Sigma(R) = f_\Sigma(0) e^{R(1/R_2 - 1/R_1)}.
\end{equation}
Following \citet{BlandHawthornGerhard2016-MilkyWayReview}, we adopt values for the thick disk scale radius $R_1=2.0$ kpc, thin disk scale radius $R_2=2.5$ kpc, and $f_\Sigma(0)=0.27$.
We note that most previous studies which use the two-infall model \citep[e.g.,][]{Chiappini1997-TwoInfall,Matteucci2006-BimodalDTDConsequences,Matteucci2009-DTDModels,Spitoni2019-TwoInfall} do not consider gas outflows and instead adjust the nucleosynthetic yields to reproduce the solar abundance. We adopt radially-dependent outflows as in \citetalias{Johnson2021-Migration} (see their Section 2.4 for details) for all our SFHs, including two-infall. We discuss the implications of this difference in Section \ref{sec:two-infall-discussion}.

\subsection{Observational Sample}
\label{sec:observational-sample}

\begin{table*}
    \centering
    \caption{Sample selection parameters and median uncertainties from APOGEE DR17 (see Section \ref{sec:observational-sample}).}
    \label{tab:sample}
    \begin{tabular}{lll}
        \hline\hline
        \multicolumn{1}{c}{Parameter} & \multicolumn{1}{c}{Range or Value} & \multicolumn{1}{c}{Notes} \\
        \hline
        $\log g$            & $1.0 < \log g < 3.8$          & Select giants only \\
        $T_{\rm eff}$       & $3500 < T_{\rm eff} < 5500$ K & Reliable temperature range \\
        $S/N$               & $S/N > 80$                    & Required for accurate stellar parameters \\
        ASPCAPFLAG Bits     & $\notin$ 23                   & Remove stars flagged as bad \\
        EXTRATARG Bits      & $\notin$ 0, 1, 2, 3, or 4     & Select main red star sample only \\
        Age                 & $\sigma_{\rm Age} < 40\%$     & Age uncertainty from \citetalias{Leung2023-Ages} \\
        $R_{\rm gal}$     & $3 < R_{\rm gal} < 15$ kpc    & Eliminate bulge \& extreme outer-disk stars \\
        $|z|$               & $|z| < 2$ kpc                 & Eliminate halo stars \\
        \hline
    \end{tabular}
\end{table*}

\begin{table*}
\centering
\caption{Number of APOGEE stars in each Galactic region.}
\label{tab:apogee-regions}
  \begin{tabular}{r|cccccc}
\hline\hline
$R_{\rm gal}\in$ & $(3, 5]$ kpc & $(5, 7]$ kpc & $(7, 9]$ kpc & $(9, 11]$ kpc & $(11, 13]$ kpc & $(13, 15]$ kpc \\
$|z|\in$ &  &  &  &  &  &  \\
\hline
$(1.0, 2.0]$ kpc & 2013 & 2100 & 8734 & 3663 & 1324 & 363 \\
$(0.5, 1.0]$ kpc & 2487 & 3490 & 13811 & 9069 & 3289 & 460 \\
$(0.0, 0.5]$ kpc & 3296 & 7029 & 17319 & 16276 & 6336 & 812 \\
\hline
\end{tabular}
\unskip\label{output/apogee_regions_table.tex}\unskip%

\end{table*}

\begin{table*}
    \centering
    \caption{Median and dispersion in APOGEE parameter uncertainties.}
    \label{tab:uncertainties}
    \begin{tabular}{lcc}
        \hline\hline
        \multicolumn{1}{c}{Parameter} & \multicolumn{1}{c}{Median Uncertainty} & \multicolumn{1}{c}{Uncertainty Dispersion ($95\%-5\%$)} \\
        \hline
        [Fe/H]          & $0.0089$   & $0.0060$ \\
        ${\rm [O/Fe]}$  & $0.019$    & $0.031$ \\
        log(Age/Gyr)    & $0.10$     & $0.16$ \\
        \hline
    \end{tabular}
\end{table*}

We compare our model results to abundance measurements from the final data release \citep[DR17;][]{Abdurro'uf2022-SDSSIV-DR17} of the Apache Point Observatory Galactic Evolution Experiment \citep[APOGEE;][]{Majewski2017-APOGEE}. APOGEE used infrared spectrographs \citep{Wilson2019-APOGEE-Spectrographs} mounted on two telescopes: the 2.5-meter Sloan Foundation Telescope \citep{Gunn2006-SloanTelescope} at Apache Point Observatory in the Northern Hemisphere, and the Ir{\'e}n{\'e}e DuPont Telescope \citep{BowenVaughan1973-DuPontTelescope} at Las Campanas Observatory in the Southern Hemisphere. After the spectra were passed through the data reduction pipeline \citep{Nidever2015-APOGEE-DataReduction}, the APOGEE Stellar Parameter and Chemical Abundance Pipeline \citep[ASPCAP;][]{Holtzmann2015-ASPCAP,GarciaPerez2016-ASPCAP} extracted chemical abundances using the model grids and interpolation method described by \citet{Jonsson2020-APOGEE-DR16}.

We restrict our sample to red giant branch and red clump stars with high-quality spectra. Table \ref{tab:sample} lists our selection criteria, which largely follow from \citet{Hayden2015-ChemicalCartography}. This produces a final sample of %
  \num{171635}\unskip\label{output/sample_size.txt}\unskip%
 stars with calibrated [O/Fe] and [Fe/H] abundance measurements. We make use of the {\it Gaia} Early Data Release 3 (EDR3) data \citep{Gaia2016-Mission,Gaia2021-EDR3} included in the catalog by the APOGEE team. Specifically, we use the \citet{Bailer-Jones2021-GaiaDistances} photo-geometric distance estimates to calculate galactocentric radius $R_{\rm gal}$ and midplane distance $z$, assuming a Sun--Galactic center distance $R_\odot=8.122$ kpc \citep{GRAVITY2018-GalactocentricDistance} and height of the Sun above the midplane $z_\odot=20.8$ pc \citep{BennetBovy2019-SunZHeight}. Table \ref{tab:apogee-regions} lists the number of APOGEE stars in bins of $R_{\rm gal}$ and $|z|$. For some Galactic regions with $R_{\rm gal}<5$ kpc or $R_{\rm gal}>13$ kpc, the median distance error exceeds 1 kpc but remains within our bin width of 2 kpc, and the vast majority of stars have much smaller distance uncertainties.

We use estimated ages from \citet[][hereafter \citetalias{Leung2023-Ages}]{Leung2023-Ages}, who use a variational encoder-decoder network which is trained on asteroseismic data to retrieve age estimates for APOGEE giants without contamination from age-abundance correlations. Importantly, the \citetalias{Leung2023-Ages} ages do not plateau beyond $\sim10$ Gyr as they do in astroNN \citep{Mackereth2019-astroNN-Ages}. We use an age uncertainty cut of 40\% per the recommendations of \citetalias{Leung2023-Ages}, which produces a total sample of %
  \num{57607}\unskip\label{output/age_sample_size.txt}\unskip%
 APOGEE stars with age estimates. We note that we use the full sample of %
  \num{171635}\unskip\label{output/sample_size.txt}\unskip%
 APOGEE stars unless we explicitly compare to age estimates. Table \ref{tab:uncertainties} presents the median and dispersion ($95^{\rm th} - 5^{\rm th}$ percentile difference) of the uncertainty in [Fe/H], [O/Fe], and log(age).

\section{One-Zone Models}
\label{sec:onezone-results}

Before running the full multi-zone models, it is useful to understand the effects of the DTD in more idealized conditions. A one-zone model assumes the entire gas reservoir is instantaneously mixed, removing all spatial dependence. This limits the ability to compare to observations across the disk, but it obviates the complicating factor of stellar migration and better isolates the effects of the nucleosynthesis prescription. In this section, we compare the results from one-zone models which examine various parameters of the DTD while keeping other parameters fixed. We use the outputs of our one-zone models to identify the regions in chemical abundance space which are most sensitive to the DTD.

For consistency, we adopt most of the parameter values from Table \ref{tab:multizone-parameters} for our one-zone models.
We adopt the inside-out SFR (Equation \ref{eq:insideout-sfh}) evaluated at $R_{\rm gal}=8$ kpc (i.e., $\tau_{\rm rise}=2$ Gyr and $\tau_{\rm sfh}=15.1$ Gyr) and an SFE timescale $\tau_\star\equiv M_g/\dot M_\star=2$ Gyr. Unless otherwise specified, we adopt an outflow mass-loading factor $\eta\equiv \dot M_{\rm out}/\dot M_\star=2.15$ \citepalias[see Equation 8 from][]{Johnson2021-Migration} and a minimum SN Ia delay time $t_D=40$ Myr. 

\subsection{DTD Parameters: Slope, Timescale, and Plateau Width}
\label{sec:onezone-dtd-params}

\begin{figure*}
    \centering
    \includegraphics[width=\linewidth]{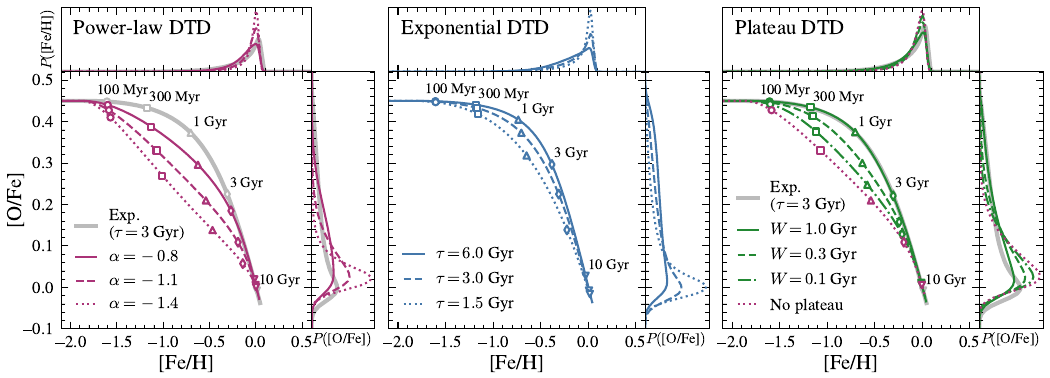}
    \caption{Abundance tracks in the [O/Fe]--[Fe/H] plane for one-zone chemical evolution models (see discussion in Section \ref{sec:onezone-results}) which assume the various DTD shapes (see Figure \ref{fig:dtds}). The open symbols along each curve mark logarithmic steps in time. The top and right-hand marginal panels present the distribution functions (DFs) of [Fe/H] and [O/Fe], respectively. For display purposes, these distributions are convolved with a Gaussian kernel with a standard deviation of 0.02 dex. 
    \textit{Left:} A power-law DTD with varying slope $\alpha$. For reference, the solid gray curve represents an exponential DTD with $\tau=3$ Gyr. 
    \textit{Center:} An exponential DTD with varying timescale $\tau$. 
    \textit{Right:} A plateau DTD with varying width $W$. All assume a post-plateau slope of $\alpha=-1.1$. For reference, the solid gray curve represents an exponential DTD with $\tau=3$ Gyr, and the dotted purple curve represents a power-law DTD with $\alpha=-1.1$ and no plateau.}
    \label{fig:onezone-threepanel}
    \script{onezone_threepanel.py}
\end{figure*}

The left-hand panel of Figure \ref{fig:onezone-threepanel} compares the results of three one-zone models that are identical except for the slope of the power-law DTD. A steeper slope implies a greater number of prompt SNe Ia which rapidly enrich the ISM with Fe, producing a faster decline in [O/Fe] with increasing [Fe/H] and hence a sharper ``knee'' after the minimum delay time. This results in a narrower distribution of [O/Fe] around the low-$\alpha$ sequence and a dearth of high-$\alpha$ stars. In all cases the [O/Fe] DF is distinctly unimodal. The MDF is not as strongly affected by the power-law slope: a shallower slope results in only a modest increase in the width of the distribution. The abundance tracks converge to the equilibrium value, reflecting the yield ratio of CCSNe to SNe Ia which is the same in all models.

Similar trends can be seen when adjusting the timescale of the exponential DTD, as shown in the middle panel of Figure \ref{fig:onezone-threepanel}. Here, the knee is not a sharp feature associated with the onset of SNe Ia as in the power-law case, but rather a gentle curve in the abundance track around $t=1$ Gyr. Doubling the timescale from 1.5 Gyr to 3 Gyr implies a longer median delay time, which raises the [O/Fe] abundance ratio at $t=1$ Gyr by $\sim0.05$ dex and at $t=3$ Gyr by $\sim0.1$ dex. A longer exponential timescale also produces a broader [O/Fe] DF with more high-$\alpha$ stars, but the distribution is still unimodal. The effect on the MDF is slightly more pronounced than the power-law case, with longer timescales skewing to lower [Fe/H] values.

Finally, the right-hand panel of Figure \ref{fig:onezone-threepanel} shows the effect of varying the width $W$ of the plateau DTD. The abundance tracks from several different plateau widths fill the space in between the exponential ($\tau=3$ Gyr) and power-law ($\alpha=-1.1$ with no plateau) models, which are both included in the panel for reference. The plateau ($W=1$ Gyr) and exponential ($\tau=3$ Gyr) DTDs produce nearly identical abundance tracks but their [O/Fe] DFs are more distinct, illustrating the need for both observables to discriminate between DTDs. The effect on the [O/Fe] DF is similar to the previous two models: a longer plateau raises the median delay time, producing a broader [O/Fe] DF and a more prominent high-$\alpha$ tail. On the other hand, all of the plateau DTDs produce very similar MDFs.

Unlike the previous three DTDs, we fix the hyper-parameters of the two-population and triple-system models to reproduce specific DTDs from the literature. A variant of the two-population DTD with a broader Gaussian component ($t_{\rm max}=0.1$ Gyr, $\sigma=0.03$ Gyr) produced similar abundance tracks and a nearly identical [O/Fe] distribution to the fiducial parametrization in a one-zone model. The effect of different parameters for the triple-system DTD would be very similar to the plateau DTD described above, of which the triple-system DTD is a special case.

\subsection{The Minimum SN Ia Delay Time}
\label{sec:onezone-minimum-delay}

\begin{figure*}
    \centering
    \includegraphics[width=\linewidth]{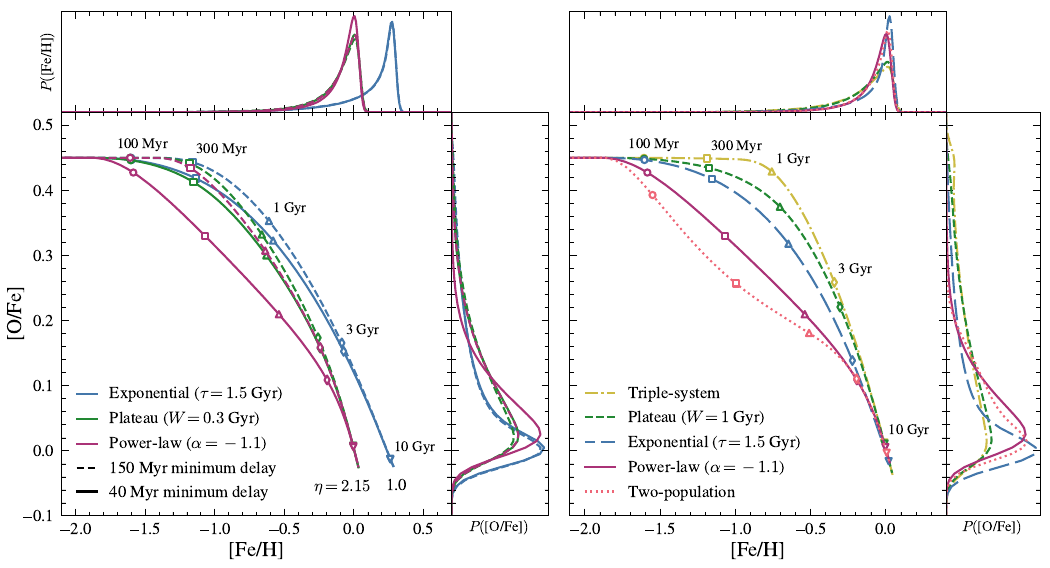}
    \caption{\textit{Left:} Comparison of one-zone models with different combinations of minimum delay time $t_D$ and DTD shape.
    The layout is similar to Figure \ref{fig:onezone-threepanel}. For visual clarity, we assume a mass-loading factor $\eta=1$ for the exponential DTD curves, which places the end-point of the abundance tracks at higher [Fe/H].
    \textit{Right:} Comparison of one-zone models with five different DTD models (see Figure \ref{fig:dtds}).
    }
    \label{fig:onezone-twopanel}
    \script{onezone_twopanel.py}
\end{figure*}

We also explore the effect of varying the minimum SN Ia delay time $t_D$ (Section \ref{sec:minimum-delay}).
The left-hand panel of Figure \ref{fig:onezone-twopanel} shows that $t_D$ has a much stronger effect in models which assume a power-law DTD than others. This is a consequence of the high number of prompt SNe Ia ($t\lesssim100$ Myr; see Figure \ref{fig:dtds}). Moreover, a power-law DTD with a long $t_D$ may be observationally hard to distinguish from a plateau model. In Figure \ref{fig:onezone-twopanel}, the abundance track for the model with a power-law DTD and $t_D=150$ Myr (dashed purple line) is similar to that of the plateau DTD with $W=0.3$ Gyr and $t_D=40$ Myr (solid green line), and their [O/Fe] DFs are virtually identical. For the exponential ($\tau=3$ Gyr) DTD, the two values of $t_D$ produce nearly indistinguishable outputs. We do not consider the effect on the other DTDs because a 150 Myr minimum delay time is incompatible with the two-population model, which has $\sim 50$\% of SNe Ia explode in the first 100 Myr, and would have a negligible effect on the triple-system DTD due to its low SN Ia rate at short delay times.
In the multi-zone models, we will hold $t_D$ fixed at 40 Myr.

\subsection{The Form of the DTD}
\label{sec:onezone-dtd-form}

The right-hand panel of Figure \ref{fig:onezone-twopanel} compares the one-zone model outputs from the full range of DTDs we investigate in this paper. As with the individual DTD parameters, the form of the DTD primarily affects the location of the high-$\alpha$ knee in the [O/Fe]--[Fe/H] abundance tracks. At one extreme is the triple-system model, which sees the CCSN plateau extend up to ${\rm [Fe/H]}\approx-0.8$ followed by a sharp downward turn as the SN Ia rate suddenly increases at a delay time of 500 Myr. 
At the other extreme are the two-population and power-law ($\alpha=-1.1$) DTDs, for which the SN Ia rate peaks immediately after the minimum delay time of 40 Myr, placing the high-$\alpha$ knee at ${\rm [Fe/H]}\approx-1.8$. The two-population model has a unique second knee at ${\rm [Fe/H]}\approx-0.2$ and ${\rm [O/Fe]}\approx0.1$, which is produced by the delayed exponential component, as noted by \citet{Vincenzo2017-ChemicalEvolution}. The abundance tracks from the plateau ($W=1$ Gyr) and exponential ($\tau=1.5$ Gyr) models occupy the intermediate space between these extremes. 

The [O/Fe] DFs also show significant differences between the DTDs. In the triple-system model, star formation proceeds for such a long time before the knee that the [O/Fe] DF shows a slight second peak around the CCSN yield ratio ($\sim0.45$ dex). Out of all our one-zone models, this small bump is the only degree of bimodality that arises in the [O/Fe] DF. Below ${\rm [O/Fe]}\approx0.4$, the plateau ($W=1$ Gyr) and triple-system DTDs produce nearly identical distributions, while the exponential DTD produces the narrowest distribution. The power-law ($\alpha=-1.1$) and two-population DTDs produce similar [O/Fe] DFs despite notably different abundance tracks. The exponential ($\tau=3$ Gyr) and plateau ($W=0.3$ Gyr) models, while not shown, produce similar abundance tracks to the plateau ($W=1$ Gyr) and exponential ($\tau=1.5$ Gyr) models, respectively.
The DTD also slightly shifts the peak of the [O/Fe] DF, with the exponential DTD placing it $\sim 0.02$ dex lower than the power-law DTD. We see similar trends in the MDF, but to a lesser degree.

The results presented in this section indicate that the [O/Fe]--[Fe/H] abundance tracks and the [O/Fe] DF are most sensitive to the parameters of the DTD, while the MDF is a less sensitive diagnostic. Degeneracies between models in one regime can be resolved in the other. For example, the exponential ($\tau=3$ Gyr) and plateau ($W=1$ Gyr) DTDs are indistinguishable in [O/Fe]--[Fe/H] space but predict different [O/Fe] DFs. Of course, both of these observables are also greatly affected by the parameters of the SFH. In this section we focused on the fiducial inside-out SFH. \citet{Palicio2023-AnalyticDTD} compared similar DTDs in one-zone models with a two-infall SFH (see Section \ref{sec:two-infall-discussion}).

\section{Multi-Zone Models}
\label{sec:multizone-results}

We use the multi-zone GCE model tools in \vice developed by \citetalias{Johnson2021-Migration}. The basic setup of our models follows theirs. The disk is divided into concentric rings of width $\delta R_{\rm gal}=100$ pc. Stellar populations migrate radially under the prescription we describe in Appendix \ref{app:migration}, but each ring is otherwise described by a conventional one-zone GCE model with instantaneous mixing (see discussion in Section \ref{sec:onezone-results}). Following \citetalias{Johnson2021-Migration}, we do not implement radial gas flows \citep[e.g.,][]{LaceyFall1985-RadialGasFlows,BilitewskiSchonrich2012-RadialFlows}. Stellar populations are also assigned a distance from the midplane according to their age and final radius as described in Appendix \ref{app:migration}.

We run our models with a time-step size of $\Delta t=10$ Myr up to a maximum time of $t_{\rm max}=13.2$ Gyr. Following \citetalias{Johnson2021-Migration}, we set \vice to form $n=8$ stellar populations per ring per time-step, and we set a maximum star-formation radius of $R_{\rm SF} = 15.5$ kpc, such that $\dot\Sigma_\star=0$ for $R_{\rm gal}>R_{\rm SF}$. The model has a full radial extent of 20 kpc, allowing a purely migrated population to arise in the outer 4.5 kpc. We adopt continuous recycling, which accounts for the time-dependent return of mass from all previous generations of stars \citep[see Equation 2 from][]{JohnsonWeinberg2020-Starbursts}. We summarize these parameters in Table \ref{tab:multizone-parameters}.

We run a total of multi-zone models with all combinations of our eight DTDs and four SFHs, for a total of 32. In the following subsections, we present the stellar abundance and age distributions from the multi-zone models and compare to APOGEE data from across the Galactic disk.

\subsection{The distribution of [Fe/H]}
\label{sec:feh-df}

\begin{figure*}
    \centering
    \includegraphics[width=\linewidth]{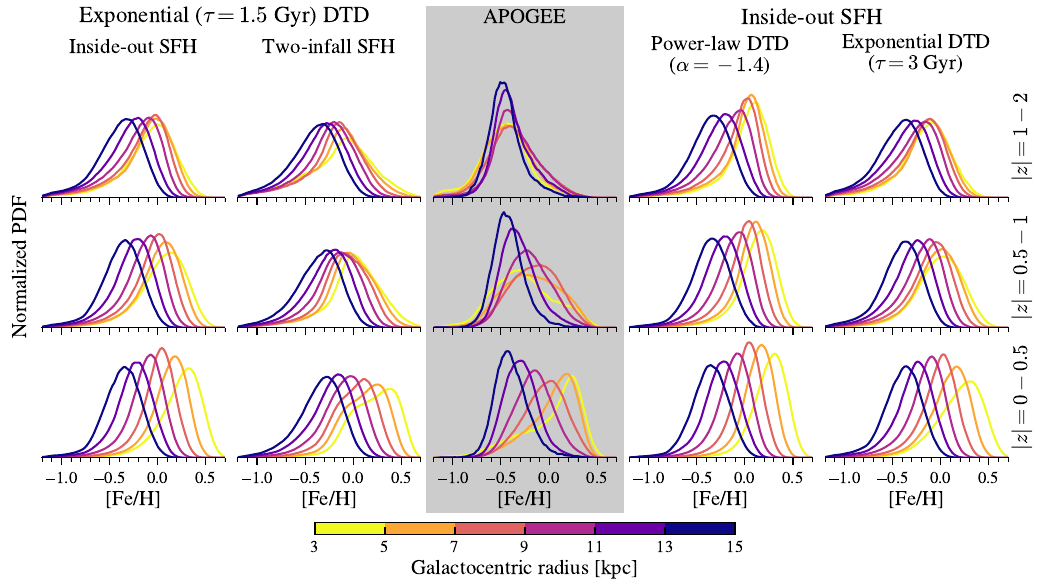}
    \caption{MDFs from multi-zone models with various SFHs and DTDs. Each row presents distributions of stars within a range of midplane distance: $1\leq|z|<2$ kpc (\textit{top}), $0.5\leq|z|<1$ kpc (\textit{middle}), and $0\leq|z|<0.5$ kpc (\textit{bottom}). Within each panel, curves of different color represent the distributions of stars binned by Galactocentric radius $R_{\rm gal}$, from $3\leq R_{\rm gal}<5$ kpc (yellow) to $13\leq R_{\rm gal}<15$ kpc (blue). Each distribution is normalized so the area under the curve is 1, and the vertical scale is consistent across each row. A Gaussian scatter with a width equal to the median observational uncertainty in APOGEE DR17 (see Table \ref{tab:uncertainties}) is applied to the abundance of each model stellar population. For visual clarity, each MDF is smoothed with a box-car width of 0.2 dex. 
    \textit{Left columns:} comparison between the inside-out and two-infall SFHs; both assume the exponential ($\tau=1.5$ Gyr) DTD. 
    \textit{Center column:} the distributions from APOGEE DR17 for reference, binned and smoothed similarly.
    \textit{Right columns:} comparison between the power-law ($\alpha=-1.4$) and exponential ($\tau=3$ Gyr) DTDs with the inside-out SFH. The MDFs in the inner Galaxy show the greatest change between the DTDs (see discussion in Section \ref{sec:feh-df}).}
    \label{fig:feh-df-comparison}
    \script{feh_df_comparison.py}
\end{figure*}

Figure \ref{fig:feh-df-comparison} shows MDFs across the Galaxy for a selection of models and APOGEE data. The two left-hand columns illustrate the effect of different SFHs on the model outputs, which is most pronounced in the inner Galaxy. Near the midplane and in the inner Galaxy, the two-infall SFH produces a distinct bump $\sim0.4$ dex below the MDF peak, which is not seen for the inside-out SFH. Away from the midplane, the low-metallicity tail is slightly more prominent for the two-infall than the inside-out model, and the two-infall MDFs extend to slightly higher metallicity. In the outer Galaxy, though, the MDFs produced by the two models are nearly identical. The shift in the skewness and peak of the MDF from the inner to the outer Galaxy is unaffected by the choice of SFH.

Holding the SFH fixed, varying the DTD has a minimal effect on the MDFs. The two right-hand columns of Figure \ref{fig:feh-df-comparison} plot the MDFs for two multi-zone models, which both assume an inside-out SFH but different DTDs: a power-law with slope $\alpha=-1.4$, and an exponential with timescale $\tau=3$ Gyr. The balance between prompt and delayed SNe Ia is starkly different between the two models, with $\sim 80\%$ of explosions occurring within 1 Gyr in the former but only $\sim 30\%$ in the latter. However, the effect on the MDF is interestingly small given this difference. The steep power-law leads to an MDF at small $R_{\rm gal}$ that is only slightly narrower than the extended exponential (made apparent by the higher peak of the normalized MDF). This tracks with our findings from one-zone models in Section \ref{sec:onezone-results} that the DTD has a smaller effect on the MDF than other observables.

The inner Galaxy MDF is more sensitive to the choice of DTD than the outer Galaxy. Here, the SFH peaks earlier and declines more sharply due to the inside-out formation of the disk. Consequently, SNe Ia often explode when the gas supply is significantly lower than when the progenitors formed. This so-called ``gas-starved ISM'' effect drives a faster increase in metallicity \citep[see analytic demonstration in][]{Weinberg2017-ChemicalEquilibrium}, which ultimately lowers the number of low-metallicity stars. The more extended the DTD, the stronger the effect. The outer disk is less affected by the choice of DTD, though, due to the more extended SFH.

To quantify the agreement between the MDFs generated by \vice and those observed in APOGEE, we compute the Kullback-Leibler (KL) divergence \citep{KullbackLeibler1951}, defined as
\begin{equation}
    D_{\rm{KL}}(P||Q) \equiv \int_{-\infty}^{\infty} p(x) \log\Big(\frac{p(x)}{q(x)}\Big) dx
    \label{eq:kl-divergence}
\end{equation}
for distributions $P$ and $Q$ with probability density functions (PDFs) $p(x)$ and $q(x)$. If $D_{\rm KL}=0$, the two distributions contain equal information. In this case, $P$ is the APOGEE MDF, $Q$ is the model MDF, and $x={\rm [Fe/H]}$. We forward-model the observational uncertainties given in Table \ref{tab:uncertainties} by applying a random Gaussian scatter to the abundance of each model stellar population, and we numerically evaluate Equation \ref{eq:kl-divergence} with integration step size $d{\rm [Fe/H]}=0.01$ dex.
For each SFH and DTD, we compute $D_{\rm{KL}}$ in the 18 different Galactic regions shown in Figure \ref{fig:feh-df-comparison}. We use bins in $R_{\rm gal}$ with a width of 2 kpc between 3 and 15 kpc, and bins in midplane distance of $|z|=0-0.5$ kpc, $0.5-1$ kpc, and $1-2$ kpc. The score $S$ for the entire model is taken to be the average of $D_{\rm{KL}}$ for each region $(R_{\rm gal}, |z|)$ weighted by the number of APOGEE stars in that region $N_\star(R_{\rm gal}, |z|)$ (see Table \ref{tab:apogee-regions}):
\begin{equation}
    S = \frac{\sum_{R_{\rm gal}, |z|} D_{\rm KL}(P||Q|R_{\rm gal}, |z|) N_\star(R_{\rm gal}, |z|)}{\sum_{R_{\rm gal}, |z|} N_\star(R_{\rm gal}, |z|)}.
    \label{eq:feh-df-score}
\end{equation}

The model combination with the best (lowest) score for the MDF is the two-infall SFH with the triple-system DTD. The choice of SFH has a larger effect on the overall score than the DTD, and the best-performing SFH is the two-infall model. However, the difference between the best-scoring model and the worst (inside-out SFH with the $\alpha=-1.4$ power-law DTD) is fairly small. While there are some {\it quantitative} differences in how the shape of the MDF varies with Galactic region, the {\it qualitative} trends are unaffected by the choice of model SFH or DTD. These trends are primarily driven by the assumption of chemical equilibrium, the abundance gradient, and radial migration (see discussion in section 3.2 of \citetalias{Johnson2021-Migration}).

\subsection{The distribution of [O/Fe]}
\label{sec:ofe-df}

\begin{figure*}
    \centering
    \includegraphics[width=\linewidth]{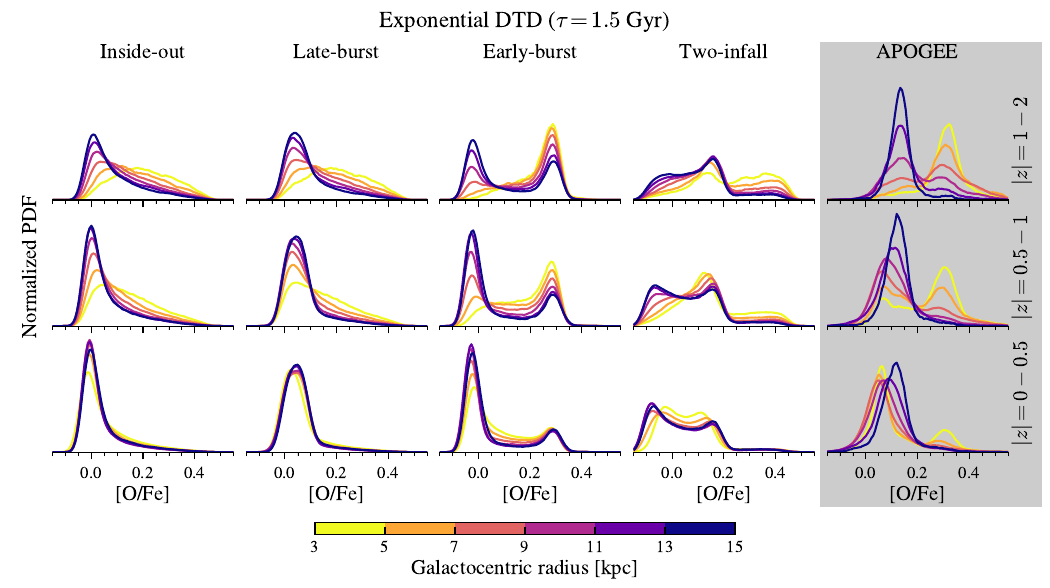}
    \caption{Distributions of [O/Fe] from multi-zone models with different SFHs. All assume the exponential ($\tau=1.5$ Gyr) DTD. The format of each panel is the same as in Figure \ref{fig:feh-df-comparison}, except that all distributions are smoothed with a box-car width of 0.05 dex. Distributions from APOGEE DR17, binned and smoothed similarly, are presented in the right-most column for reference.}
    \label{fig:ofe-df-sfh}
    \script{ofe_df_sfh.py}
\end{figure*}

The distribution of [O/Fe] serves as a record of the relative rates of SNe Ia and CCSNe. As such, its shape is affected by both the SFH and DTD. Figure \ref{fig:ofe-df-sfh} shows the distribution of [O/Fe] across the disk for the four model SFHs compared to the distributions measured by APOGEE. All four models assume an exponential DTD with $\tau=1.5$ Gyr, which has an intermediate median delay time among all our DTDs. We see similar trends with Galactic region across all four models. Near the midplane, the distributions depend minimally on radius, but away from the midplane, there is a clear trend toward higher [O/Fe] at small $R_{\rm gal}$.

While trends with $R_{\rm gal}$ and $|z|$ are similar across the different models, the shape of the distribution varies greatly with the chosen SFH. The inside-out and late-burst models produce similar distributions because of the similarity of their underlying SFHs, as the burst is imposed upon the inside-out SFH (see Equation \ref{eq:lateburst-sfh}). Both skew heavily toward near-solar [O/Fe], although the late-burst model produces a slightly broader peak and a less-prominent high-[O/Fe] tail. This difference arises because the late-burst SFH shifts a portion of the stellar mass budget to late times when [O/Fe] is low. The only region which shows any significant skew toward high [O/Fe] is $R_{\rm gal}=3-5$ kpc and $|z|=1-2$ kpc, but the shift to higher [O/Fe] at high latitudes is gradual and does not produce the notable trough at ${\rm [O/Fe]}\approx0.2$ which is seen in the APOGEE data. 

On the other hand, the early-burst model produces a bimodal [O/Fe] distribution in most regions. Although agreement is not perfect, the early-burst SFH produces the closest match to the data by far. In particular, the low-$\alpha$ sequence away from the midplane is dominated by stars in the solar annulus and outer disk, a trend which is also seen in APOGEE. However, the early-burst high-$\alpha$ sequence contains many stars in the outer disk and close to the midplane, whereas the APOGEE distribution does not show a prominent high-$\alpha$ peak beyond $R_{\rm gal}\sim11$ kpc ($\sim7$ kpc in the midplane). 

The two-infall SFH produces \textit{three} distinct modes at ${\rm [O/Fe]}\approx -0.05$, $0.15$, and $0.4$. At small $R_{\rm gal}$ and with increasing $|z|$, the low-$\alpha$ peak decreases in prominence as the high-$\alpha$ peak increases, but the intermediate peak is a striking feature at all latitudes that does not align with observations. In the APOGEE data, the high-$\alpha$ peak is at ${\rm [O/Fe]}\approx0.3$, roughly halfway between the intermediate and high-$\alpha$ peaks produced by the two-infall model. However, the model high-$\alpha$ sequence does match the observed trends with $R_{\rm gal}$ and $|z|$ much better than the early-burst models.

\begin{figure*}
    \centering
    \includegraphics[width=\linewidth]{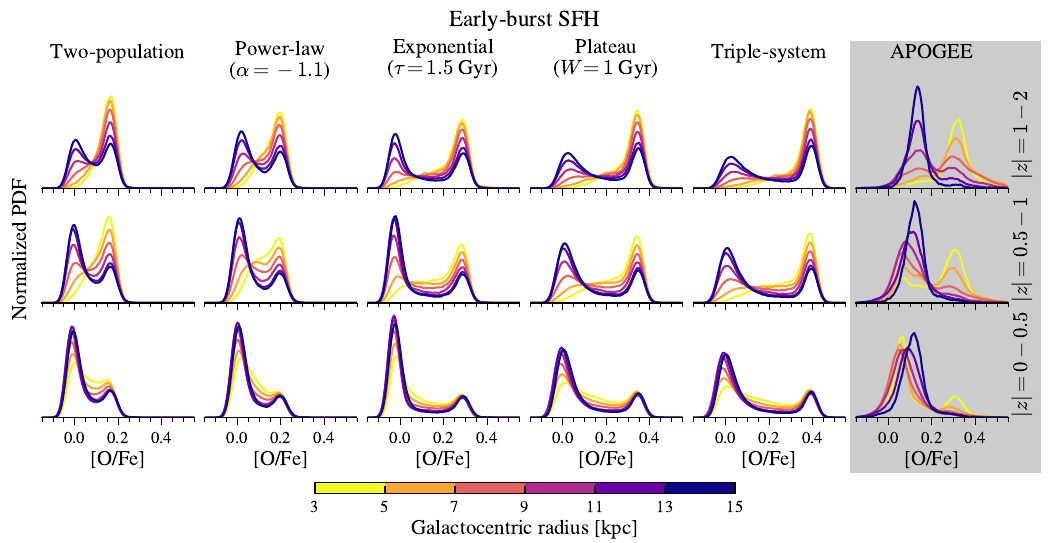}
    \caption{The same as Figure \ref{fig:ofe-df-sfh} but for different DTDs. In all cases an early-burst SFH is assumed.}
    \label{fig:ofe-df-dtd}
    \script{ofe_df_dtd.py}
\end{figure*}

Figure \ref{fig:ofe-df-dtd} shows [O/Fe] distributions produced by models with the same SFH but a range of different DTDs. We show models with the early-burst SFH because it produces distinct low- and high-$\alpha$ sequences. The most obvious effect of the DTD is to shift the mode of the high-$\alpha$ sequence. The two-population DTD, which has the most prompt SNe Ia, places the high-$\alpha$ sequence at ${\rm [O/Fe]}\approx 0.15$, while the triple-system DTD, which has the fewest prompt SNe Ia, places it $\sim0.25$ dex higher at ${\rm [O/Fe]}\approx0.4$. The plateau ($W=1$ Gyr) DTD places the higher peak at ${\rm [O/Fe]}\approx0.35$, close to where it appears in the APOGEE distributions. However, the distance {\it between} the peaks of the APOGEE distributions is only $\sim0.2$ dex, since the observed low-$\alpha$ sequence sits at ${\rm [O/Fe]}\approx0.1$. This spacing is best replicated by the power-law ($\alpha=-1.1$) DTD, even though both peaks sit $\sim0.1$ dex too low and the distributions are narrower than observed. 

In general, models with fewer prompt SNe Ia populate the high-$\alpha$ sequence with more stars because the chemical evolution track spends more time in the high-$\alpha$ regime. This qualitatively agrees with the isolated and cosmological simulations of \citet{Poulhazan2018-PrecisionPollution}, who find that DTDs with a significant prompt component produce narrower [O/Fe] distributions and a higher average [O/Fe].

We again compute the KL divergence (Equation \ref{eq:kl-divergence}) to quantify the agreement between the [O/Fe] DFs of our models and APOGEE. We calculate a score for each model as described in Section \ref{sec:feh-df}. The best-scoring model combines the inside-out SFH with the triple-system DTD, and the plateau ($W=1$ Gyr) and exponential ($\tau=3$ Gyr) DTDs score well when combined with either the inside-out or late-burst SFHs. Both plateau DTDs also score relatively well with the two-infall SFH. Surprisingly, the early-burst SFH scores quite poorly for all DTD models, despite the fact that it produces the most distinct high- and low-$\alpha$ sequences. We discuss this further in Section \ref{sec:discussion-scores}.

\subsection{Bimodality in [O/Fe]}
\label{sec:bimodality}

\begin{figure*}
    \centering
    \includegraphics[width=\linewidth]{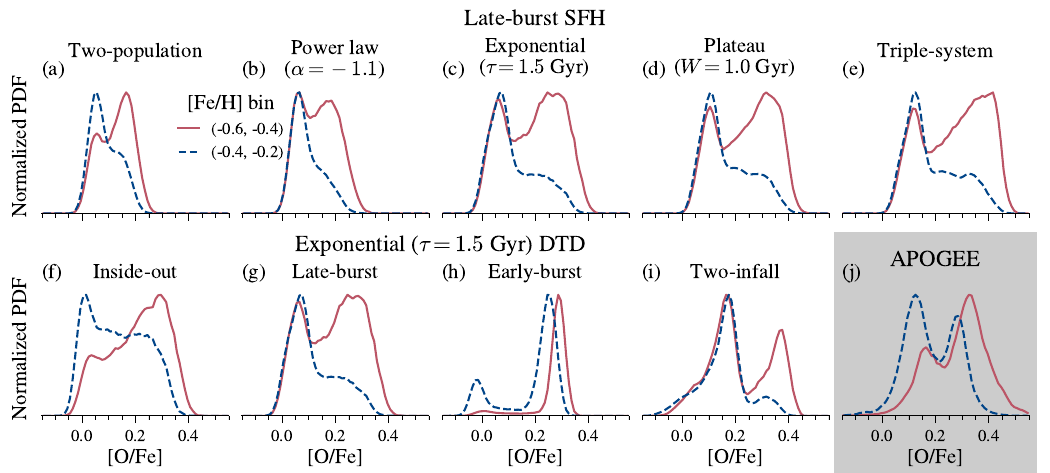}
    \caption{The distributions of [O/Fe] along two different slices of [Fe/H]: $-0.6\leq$[Fe/H]$<-0.4$ (red solid) and $-0.4\leq$[Fe/H]$<-0.2$ (blue dashed). Each panel contains stars within the Galactic region defined by $7\leq R_{\rm gal}<9$ kpc and $0\leq|z|<2$ kpc. For each distribution, \num{100000} stellar populations are re-sampled from the model output to match the $|z|$ distribution of the APOGEE sample.
    \textit{Top row:} results from five multi-zone models which assume the late-bust SFH but different DTDs. \textit{Bottom row}: the first four panels compare the four SFHs (see Figure \ref{fig:sfhs}), all assuming an exponential DTD with $\tau=1.5$ Gyr. The bottom-right panel (highlighted) plots data from APOGEE DR17 for reference.}
    \label{fig:ofe-bimodality}
    \script{ofe_bimodality_summary.py}
\end{figure*}

The [O/Fe] distributions from APOGEE in Section \ref{sec:ofe-df} show two distinct peaks whose relative prominence varies with $R_{\rm gal}$ and $|z|$ \citep[see also Figure 4 of][]{Hayden2015-ChemicalCartography}. A crucial feature of this bimodality, which is not apparent in the analysis of the previous section, is the presence of both sequences at fixed [Fe/H]. The separation between the two sequences appears to be a real feature and not an artifact of the APOGEE selection function \citep{Vincenzo2021-AlphaDistribution}. A successful model for the evolution of the Milky Way therefore must reproduce this bimodality.

Figure \ref{fig:ofe-bimodality} compares the [O/Fe] distributions in the solar annulus ($7\leq R_{\rm gal}<9$ kpc and $0\leq|z|<2$ kpc) in two bins of [Fe/H] ($-0.6<{\rm [Fe/H]}<-0.4$ and $-0.4<{\rm [Fe/H]}<-0.2$) for select model outputs and APOGEE data. The purpose of the narrow [Fe/H] bins is to isolate the bimodality of the [O/Fe] distribution with minimal variation in [Fe/H]. The APOGEE distributions in the bottom-right panel (j) show that the high-$\alpha$ mode is more prominent at lower [Fe/H], but the distributions in both bins are clearly bimodal. The ``trough'' occurs near ${\rm [O/Fe]}\approx0.2$ in each bin.

To quantify the strength of the $\alpha$-bimodality, we use the peak-finding algorithm {\tt scipy.signal.find\_peaks} \citep{2020SciPy-NMeth}. For each peak, we calculate the prominence, or the vertical distance between a peak and its highest neighboring trough. We consider a distribution bimodal if both peaks exceed an arbitrary threshold of 0.1. The APOGEE distributions exceed this threshold in both [Fe/H] bins.

The top row of panels (a--e) in Figure \ref{fig:ofe-bimodality} shows the [O/Fe] bimodality (or lack thereof) across five different DTDs, all of which assume the late-burst SFH. To better approximate the APOGEE selection function, we re-sample our model stellar populations so the $|z|$ distribution closely matches that of APOGEE in the solar neighborhood. Six of the eight DTDs (all except the two-infall and $\alpha=-1.4$ power-law DTDs) exceed our prominence threshold in the low-[Fe/H] bin. Panel (a) shows that the two-infall DTD produces a marginal low-$\alpha$ peak, although it does not meet the prominence threshold. In general, DTDs with fewer prompt SNe Ia produce a high-$\alpha$ peak which is more prominent and at a higher [O/Fe], as was the case with the [O/Fe] distributions in Section \ref{sec:ofe-df}. 

Panels (f)--(i) in the bottom row of Figure \ref{fig:ofe-bimodality} illustrate the effect of the SFH on the [O/Fe] bimodality. The inside-out SFH does not produce a bimodal distribution for most of our DTDs (the exception is the $W=1$ Gyr plateau DTD, which produces a much smaller trough than observed). On the other hand, the early-burst SFH {\it always} produces a bimodal distribution in the high-[Fe/H] bin regardless of the assumed DTD, but not in the low-[Fe/H] bin (the small low-$\alpha$ peak falls below our prominence threshold). For models with the late-burst and two-infall SFHs, the bimodality in the low-[Fe/H] bin is variable depending on the DTD: those with longer median delay times (e.g., exponential, plateau, or triple-system) generally produce a bimodal distribution, while the two DTDs with the most prompt SNe Ia do not. 

One major problem in all of our models is the presence of the \aFe bimodality across only a narrow range of [Fe/H]. Even our most successful models can produce a bimodal [O/Fe] distribution in only one bin: the high-[Fe/H] bin for the early-burst SFH, and the low-[Fe/H] bin for the late-burst and two-infall SFHs. In APOGEE, the two sequences are co-extant between ${\rm [Fe/H]}\approx-0.6$, below which the high-$\alpha$ sequence dominates, and ${\rm [Fe/H]}\approx+0.2$, at which point they join. The failure of these models to fully reproduce the bimodality across the whole range of [Fe/H] was noted by \citetalias{Johnson2021-Migration}, and the problem persists for each model we consider here.

\subsection{The [O/Fe]--[Fe/H] Plane}
\label{sec:ofe-feh}

\begin{figure}
    \centering
    \includegraphics{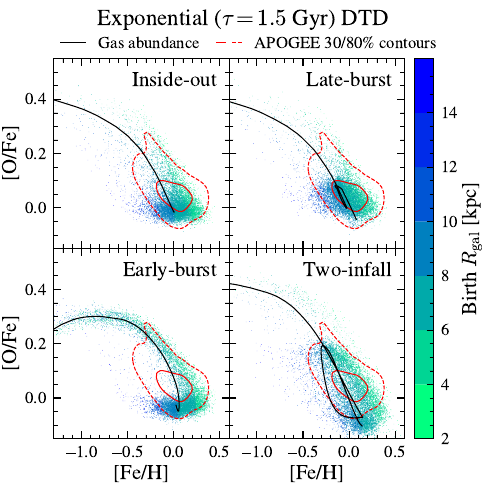}
    \caption{A comparison of the [O/Fe]--[Fe/H] plane between the four SFHs in our multi-zone models. All assume the exponential ($\tau=1.5$ Gyr) DTD. Each panel plots a random mass-weighted sample of \num{10000} star particles in the solar neighborhood ($7\leq R_{\rm gal}<9$ kpc, $0\leq|z|<0.5$ kpc) color-coded by $R_{\rm gal}$ at birth. A Gaussian scatter has been applied to all points based on the median abundance errors in APOGEE DR17 (see Table \ref{tab:sample}). The black curves represent the ISM abundance tracks in the 8 kpc zone. The red contours represent a 2-D Gaussian kernel density estimate of the APOGEE abundance distribution in that Galactic region with a bandwidth of 0.03. The solid and dashed contours enclose 30\% and 80\% of stars in the sample, respectively.}
    \label{fig:ofe-feh-sfh}
    \script{ofe_feh_sfh.py}
\end{figure}

In Section \ref{sec:onezone-results}, we illustrated that the form and parameters of the DTD have an important effect on the ISM abundance tracks in idealized one-zone models (see Figures \ref{fig:onezone-threepanel} and \ref{fig:onezone-twopanel}). However, comparisons to data are limited because the tracks neither record the number of stars that formed at each abundance, nor incorporate the effect of stellar migration. Here, we present the distribution of stellar abundances in the [O/Fe]--[Fe/H] plane alongside the ISM abundance tracks from our multi-zone models. We compare our model outputs to the observed distributions from APOGEE across the Milky Way disk.

Figure \ref{fig:ofe-feh-sfh} compares the [O/Fe]--[Fe/H] plane in the solar neighborhood ($7\leq R_{\rm gal}<9$ kpc, $0\leq|z|<0.5$ kpc) between our four model SFHs. The black curves represent the ISM abundance as a function of time in the $R_{\rm gal}=8.0-8.1$ kpc zone; in the absence of radial migration, all model stellar populations would lie close to these lines. Stellar populations to the left of the abundance tracks were born in the outer disk, while those to the right were born in the inner disk, as illustrated by the color-coding in the figure. Much of the scatter in [Fe/H] in a given Galactic region can be attributed to radial migration \citep{Edvardsson1993-ChemicalEvolution}.

The tracks predicted by all four SFHs initially follow a similar path of decreasing [O/Fe] with increasing [Fe/H]. The ISM abundance ratios of the inside-out model change monotonically over the entire disk lifetime. The stellar abundance distribution at both low- and high-[O/Fe] is composed of stars with a wide range of birth $R_{\rm gal}$. 

The late-burst model produces similar results to the inside-out model up to ${\rm [Fe/H]}\approx-0.2$ due to their similar SFHs. The Gaussian burst in its SFH introduces a loop in the ISM abundance track, as an uptick in star formation at $t\approx11$ Gyr raises the CCSN rate, leading to a slight increase in [O/Fe] before the subsequent increase in the SN Ia rate lowers the [O/Fe] once again \citep[see e.g. Figure 1 of][]{JohnsonWeinberg2020-Starbursts}. This loop slightly broadens the low-[O/Fe] stellar distribution as we observed in Section \ref{sec:ofe-df}.

This same pattern is seen much more strongly in the abundance tracks for the two-infall model. Here, the significant infall of pristine gas at $t=4$ Gyr leads to rapid dilution of the metallicity of the ISM, followed by a large burst in the SFR, which raises [O/Fe] by $\sim 0.2$ dex. We observe a ridge in the stellar abundance distribution at the turn-over point (${\rm [O/Fe]}\approx0.15$) associated with SNe Ia whose progenitors formed during the burst. This ridge roughly coincides with the upper limit of the APOGEE distribution near the midplane. The three-peaked structure of the [O/Fe] distributions in Section \ref{sec:ofe-df} is explained by the abundance tracks here: a small population of stellar populations at ${\rm [O/Fe]}\approx0.4$ is produced initially, followed by the middle peak when the abundance track turns over, and finally the peak at ${\rm [O/Fe]}\approx-0.1$ which reflects the equilibrium abundance ratio of the second infall.

The early-burst track is the most distinct from the other models at low metallicity. The portion shown in Figure \ref{fig:ofe-feh-sfh} represents the evolution {\it after} the early SFE burst. At low metallicity, there is a ``simmering phase'' where [O/Fe] slowly decreases to a local minimum at ${\rm [Fe/H]}\approx-1.3$, at which point the rapid increase in the SFE causes the [O/Fe] to rebound \citep[a more thorough examination of this behavior can be found in][]{Conroy2022-ThickDisk}.
The early-burst SFH produces the clearest separation between a high- and low-[O/Fe] sequences. The number of stars on the high-[O/Fe] sequence is relatively high, likely as a result of its higher SFR at early times compared to the other models.

\begin{figure*}
    \centering
    \includegraphics[width=\linewidth]{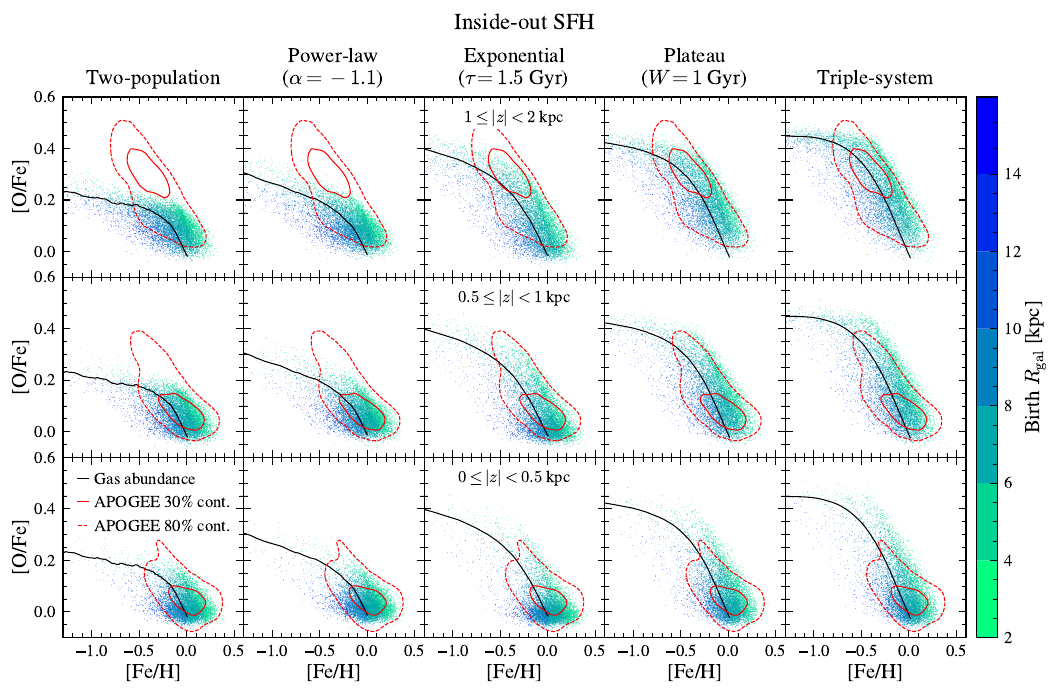}
    \caption{The [O/Fe]--[Fe/H] plane from multi-zone models with different DTDs (see Figure \ref{fig:dtds}). All assume the inside-out SFH. Each panel is similar to those in Figure \ref{fig:ofe-feh-sfh}, except each row contains star particles from a different bin in $|z|$, with stars closest to the midplane in the bottom row and stars farthest from the midplane in the top row as labeled in the middle column. All panels contain stars within the solar annulus ($7\leq R_{\rm gal}<9$ kpc).}
    \label{fig:ofe-feh-dtd}
    \script{ofe_feh_dtd.py}
\end{figure*}

Figure \ref{fig:ofe-feh-dtd} compares the [O/Fe]--[Fe/H] ISM tracks and stellar distributions for five models with the same SFH but different DTDs. We choose the inside-out SFH for this figure because it predicts monotonically-decreasing abundance ratios, making comparisons between the different DTDs relatively straightforward. The models are arranged according to the median delay time of the DTD, increasing across the panel columns from left to right. 

The two-population and power-law ($\alpha=-1.1$) DTDs, which have a large fraction of prompt ($t\lesssim100$ Myr) SNe Ia, produce stellar abundance distributions that are reasonably well-aligned with the APOGEE contours at low $|z|$, but they entirely miss the observed high-$\alpha$ sequence at large $|z|$. The ISM abundance tracks for the 8 kpc zone do not pass through the APOGEE 30\% contour at $|z|=1-2$ kpc. For both DTDs, the high-[O/Fe] knee is located below the left-most bound of the plot, but we observe a second knee at ${\rm [O/Fe]}\approx0.15$ where the abundance tracks turn downward once more. As discussed in Section \ref{sec:onezone-results}, the second knee is most prominent in the model with the two-population DTD because of its long exponential tail.

The exponential ($\tau=1.5$ Gyr) DTD, which has an intermediate median delay time, produces a distribution in Figure \ref{fig:ofe-feh-dtd} which aligns quite well with the 80\% APOGEE contours in all $|z|$-bins, and even produces a ``ridge'' which extends to high [O/Fe] at low- and mid-latitudes (bottom and center panels, respectively). While it does better at populating the high-$\alpha$ sequence than the previous DTDs, the bulk of the model stellar populations at large $|z|$ still fall below the APOGEE 30\% contour. 

The two right-hand columns present model results for the plateau ($W=1$ Gyr) and triple-system DTDs, which have the longest median delay times. The high-[O/Fe] knee occurs at a much higher metallicity in these models and is visible in the gas abundance tracks in the upper-left corner of the panels. At large $|z|$, the predicted abundance distributions align quite well with the APOGEE high-$\alpha$ sequence, but there is a significant ridge of high-$\alpha$ stars from the inner Galaxy at low $|z|$.

To quantify the agreement between the multi-zone model outputs and data in [O/Fe]--[Fe/H] space, we implement the method of \citet{PerezCruz2008-KLTest2D} for estimating the KL divergence between two continuous, multivariate samples using a $k$-nearest neighbor estimate. For $n$ samples from a multivariate PDF $p(\mathbf{x})$ and $m$ samples from $q(\mathbf{x})$, we can estimate $D_{\rm KL}(P||Q)$ according to the following:
\begin{equation}
    \label{eq:2d-kl-divergence}
    \hat D_k(P||Q) = \frac dn \sum_{i=1}^n\log\frac{r_k(\mathbf{x}_i)}{s_k(\mathbf{x}_i)} + \log\frac{m}{n-1},
\end{equation}
where $d=2$ is the dimension of the sample space and $r_k(\mathbf{x}_i)$ and $s_k(\mathbf{x}_i)$ are the distance to the $k$th nearest neighbor of $\mathbf{x}_i$ in the samples of $P$ and $Q$, respectively. We take $k=2$ to find the nearest neighbor other than the sample itself. As before, $P$ is the APOGEE distribution and $Q$ is the model distribution, and in this case $\mathbf{x}=({\rm [Fe/H]}, {\rm [O/Fe]})$, without applying any scaling factor to each dimension.
As in Sections \ref{sec:feh-df} and \ref{sec:ofe-df}, we bin the model outputs and data by $R_{\rm gal}$ and $|z|$, calculate $\hat D_k(P||Q)$ in each region, and then take the weighted mean of each region as in Equation \ref{eq:feh-df-score} to arrive at a single score for each model.

The best-scoring model combines the triple-system DTD with the inside-out SFH. The two other DTDs with the longest median delay times, plateau ($W=1$ Gyr) and exponential ($\tau=3$ Gyr), also score quite well. As with the [O/Fe] DFs, the inside-out and late-burst SFH models score similarly across all DTDs in the [O/Fe]--[Fe/H] plane. The early-burst SFH models score the worst out of all the SFHs, likely due to the long ``tail'' in the distribution down to low [Fe/H] which is not seen in the APOGEE data.

\begin{figure}
    \centering
    \includegraphics{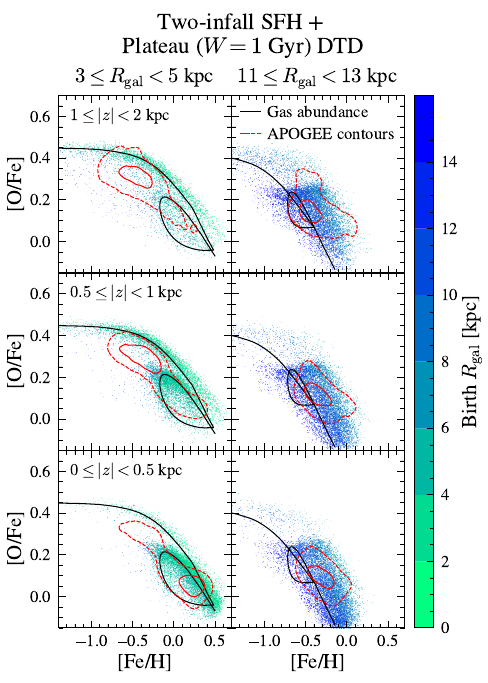}
    \caption{The [O/Fe]--[Fe/H] plane for multiple Galactocentric regions from the model with the two-infall SFH and plateau ($W=1$ Gyr) DTD. The two columns of panels contain stars in different bins of $R_{\rm gal}$, and each row contains stars from a different bin of $|z|$. The contents of each panel are as described in Figure \ref{fig:ofe-feh-sfh}.}
    \label{fig:ofe-feh-twoinfall}
    \script{ofe_feh_twoinfall.py}
\end{figure}

Figure \ref{fig:ofe-feh-twoinfall} plots the stellar [O/Fe]--[Fe/H] abundances from the model with the two-infall SFH and plateau ($W=1$ Gyr) DTD in two different bins of $R_{\rm gal}$. In the inner Galaxy, the model distribution at large $|z|$ lies at higher [O/Fe] and is more extended than the APOGEE distribution. Agreement between the model and data is worst at mid-latitudes: the model distribution is sparsest in the area of the peak of the APOGEE distribution. Near the midplane, however, the model output is well-aligned with the data. In the outer Galaxy, the distributions are well-aligned at all $|z|$, though the model distributions are more extended along the [O/Fe] axis than in the data. Adjustments to the yields or the relative infall strengths could improve the agreement between the two-infall model output and the observed distributions.

\subsection{The Age--[O/Fe] Plane}
\label{sec:age-ofe}

\begin{figure}
    \centering
    \includegraphics{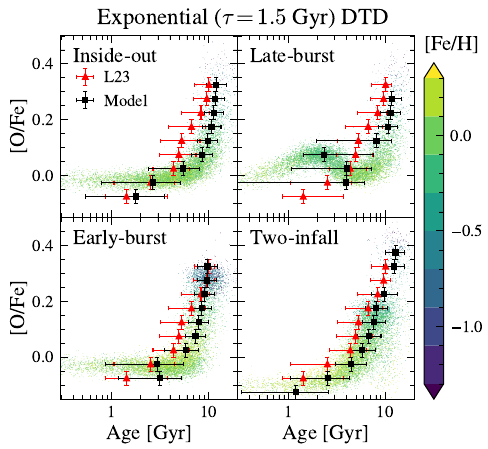}
    \caption{A comparison of the age--[O/Fe] relation between multi-zone models with different SFHs. All assume the exponential ($\tau=1.5$ Gyr) DTD. Each panel plots a random mass-weighted sample of 10,000 star particles in the solar neighborhood ($7\leq R_{\rm gal}<9$ kpc, $0\leq|z|<0.5$ kpc) color-coded by [Fe/H]. A Gaussian scatter has been applied to all points based on the median [O/Fe] error from APOGEE DR17 and the median age error from \citetalias{Leung2023-Ages} (see Table \ref{tab:sample}). Black squares represent the mass-weighted median age of star particles within bins of [O/Fe] with a width of 0.05 dex, and the horizontal black error bars encompass the 16th and 84th percentiles. Red triangles and horizontal error bars represent the median, 16th, and 84th percentiles of age from \citetalias{Leung2023-Ages}, respectively. For clarity, bins which contain less than 1\% of the total mass (in the models) or total number of stars (in the data) are not plotted.}
    \label{fig:age-ofe-sfh}
    \script{age_ofe_sfh.py}
\end{figure}

As demonstrated by our one-zone models in Section \ref{sec:onezone-results}, models that produce similar tracks in abundance space can be distinguished by the rate of their abundance evolution. We therefore expect the age--[O/Fe] relation to be a useful diagnostic. Figure \ref{fig:age-ofe-sfh} shows the stellar age and [O/Fe] distributions in the solar neighborhood for each of our four SFHs. As in Figure \ref{fig:ofe-feh-sfh}, all four panels assume an exponential DTD with $\tau=1.5$ Gyr. We compare these predictions against ages estimated with \citetalias{Leung2023-Ages}'s variational encoder-decoder algorithm. We caution against drawing strong conclusions from this comparison, because we do not correct for selection effects or systematic errors in the age determination. 

The inside-out and late-burst models show fair agreement with the data at high [O/Fe], although both show a $\sim2$ Gyr offset. One could shift the ramp-up in star formation to slightly later times or simply run the model for a shorter amount of time to close this gap. Although it is a visually striking difference, the age at the high-[O/Fe] knee is not a good diagnostic for the SFH after factoring in the age uncertainties. As in the data, the trend in the median age with decreasing [O/Fe] decreases monotonically in the inside-out model. The late-burst model, however, shows a bump in the relation at a lookback time of $\sim2$ Gyr which is not seen in the data, as noted by \citetalias{Johnson2021-Migration}.

For the early-burst SFH, the predicted stellar ages are almost perfectly aligned with the data for ${\rm [O/Fe]}\gtrsim 0.2$. The rapid rise in the SFE at early times delays the descent to lower [O/Fe] values and produces a clump of low-metallicity, high-[O/Fe] stars at an age of $\sim10$ Gyr. 
Lastly, the two-infall SFH produces a fair match to the data. Stars with ${\rm [O/Fe]}\gtrsim 0.25$ were produced in the first infall, while the second infall produces a clump of stars with similar metallicity, ages of $\sim8$ Gyr, and ${\rm [O/Fe]}\approx0.2$. There is a population of old, low-$\alpha$ stars that arise due to the initial descent in [O/Fe] prior to the second accretion epoch. The subsequent increase in [O/Fe] does not produce as strong of a bump as the late-burst SFH, because it occurs much earlier and is therefore narrower in log(age). However, the two-infall SFH produces [O/Fe] abundances for the youngest stars which are roughly 0.1 dex lower than the other models.

In contrast to \citetalias{Johnson2021-Migration}, none of our models predict a population of young, $\alpha$-enhanced stars in the solar neighborhood. These stars have been observed in APOGEE \citep[e.g.,][]{Martig2016-CNAbundances,SilvaAguirre2018-YoungAlphaEnhanced} and many are likely old systems masquerading as young stars due to mass transfer or a merger \citep[e.g.,][]{Yong2016-YoungAlphaRich}, but it is not known whether some fraction are truly intrinsically young \citep{HekkerJohnson2019-YoungAlphaRich}. In \citetalias{Johnson2021-Migration}, these young, $\alpha$-enhanced stars are the result of a highly variable SN Ia rate in the outer Galaxy. The SN Ia progenitors migrate before they are able to enrich their birth annulus, so the subsequent stellar populations are depleted in Fe. Two differences in the migration scheme explain the lack of these stars in our own models: first, we adopt a time-dependence for radial migration of $\Delta t^{1/3}$, which is slower than the diffusion scheme ($\Delta t^{1/2}$) of \citetalias{Johnson2021-Migration}. Second, our migration method is designed to produce smooth abundance distributions, whereas the method of \citetalias{Johnson2021-Migration} can assign identical migration patterns to many stellar populations in sparsely-populated regions of the Galaxy, potentially removing many SN Ia progenitors from a given zone simultaneously (for more discussion, see Appendix \ref{app:migration}). This update to the model is consistent with \citeauthor{Grisoni2024-YoungAlphaRich}'s \citeyearpar{Grisoni2024-YoungAlphaRich} finding that young $\alpha$-rich stars have similar occurrence rates across the disk, which supports a stellar, as opposed to Galactic, origin.

\begin{figure*}
    \centering
    \includegraphics[width=\linewidth]{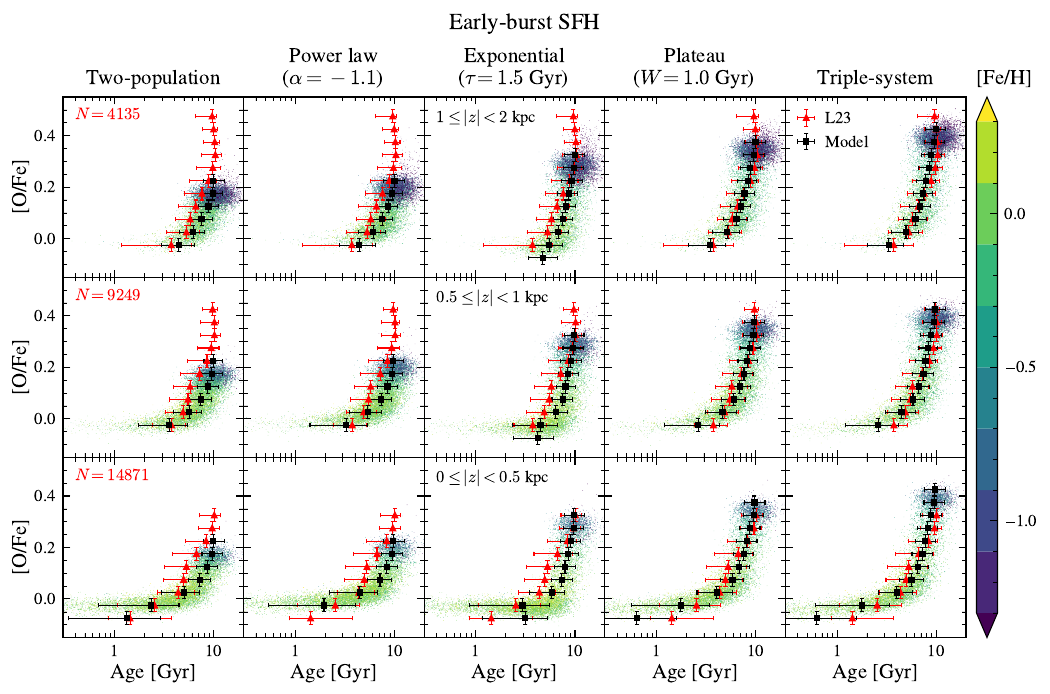}
    \caption{A comparison of the age--[O/Fe] relation between multi-zone models with different DTDs. All assume the early-burst SFH. Each row contains star particles from a different bin in $|z|$, with stars closest to the midplane in the bottom row and stars farthest from the midplane in the top row as labeled in the middle column. In all panels stars are limited to the solar annulus ($7\leq R_{\rm gal}<9$ kpc), and the layout of each panel is the same as in Figure \ref{fig:age-ofe-sfh}. The red numbers in the left-most panels indicate the number of APOGEE stars with \citetalias{Leung2023-Ages} ages in each bin of $|z|$.}
    \label{fig:age-ofe-dtd}
    \script{age_ofe_dtd.py}
\end{figure*}

Figure \ref{fig:age-ofe-dtd} shows the predicted age--[O/Fe] relation for five of our DTDs. All models were run with the early-burst SFH because it predicts the clearest separation between the high- and low-$\alpha$ sequences (see Figure \ref{fig:age-ofe-sfh}). Similar to Figure \ref{fig:ofe-feh-dtd}, models are arranged from left to right by increasing median SN Ia delay time. The high-$\alpha$ sequence moves to higher [O/Fe] with increasing median delay time, from $\sim0.2$ for the two-population model to $\sim0.4$ for the triple-system DTD. As we have seen in previous figures, the range in [O/Fe] produced by DTD models with many prompt SNe Ia is much smaller than the extended DTDs. At high $|z|$ (top row), the observed range of [O/Fe] is larger than what is produced by most of our models. While the plateau ($W=1$ Gyr) and triple-system models come close, the other three fall short of the observed range in [O/Fe], but still closely match the median age--[O/Fe] relation. There is a slight reversal in the observed trend for the stars with the highest [O/Fe]: the $0.45\leq{\rm [O/Fe]}<0.5$ bin has a slightly \textit{lower} median age than the $0.3\leq{\rm [O/Fe]}<0.35$ bin at high $|z|$ in the \citetalias{Leung2023-Ages} sample, a small effect but one which is not predicted by any of our models.

Moving to stars at low $|z|$, the plateau ($W=1$ Gyr) and triple-system DTDs over-produce stars at the old, high-$\alpha$ end of the distribution, while also diverging somewhat from the observed sequence near solar [O/Fe]. The exponential ($\tau=1.5$ Gyr) DTD comes closest to reproducing the observed range in [O/Fe], while the two DTDs with the shortest median delay time once again produce a smaller range of [O/Fe] than observed. We note that the break between the linear and flat parts of the relation is sharpest for the exponential DTD, and a more gradual transition is observed for the other four DTDs. This difference arises because the exponential DTD is most dominant at intermediate delay times ($t\sim 1-3$ Gyr) but falls off much faster than the other models at long delay times, so [O/Fe] is close to constant for lookback times $\lesssim 5$ Gyr. Overall, the exponential ($\tau=1.5$ Gyr) DTD most closely matches the data for stars with $0\leq|z|<0.5$ kpc.

We use a different scoring system from previous sub-sections due to the much larger uncertainties in age than [O/Fe]. As shown in Figures \ref{fig:age-ofe-sfh} and \ref{fig:age-ofe-dtd}, in each Galactic region we sort the model outputs and data into bins of [O/Fe] with a width of 0.05 dex. We define the root mean square (RMS) median age difference for the region as
\begin{equation}
    \Delta\tau_{\rm RMS} \equiv \sqrt{\frac{\sum_k \Delta\tau_k^2 n_{{\rm L23},k}}{n_{\rm L23,tot}}}
    \label{eq:age-ofe-score}
\end{equation}
where $\Delta\tau_k=\rm{med}(\tau_{\rm \vice})-\rm{med}(\tau_{\rm L23})$ is the difference between the mass-weighted median age in \vice and the median stellar age from \citetalias{Leung2023-Ages} in bin $k$, $n_{{\rm L23},k}$ is the number of stars from the \citetalias{Leung2023-Ages} age sample in bin $k$, and $n_{\rm L23,tot}$ is the total number of stars in the sample in that Galactic region. This is similar to a reduced $\chi^2$ estimator except that the difference in medians is not weighted by the variance in the observed sample. If a bin has no modeled or observed stars, we do not calculate $\Delta \tau_k$ for that bin. As before, the score for the model as a whole is the average of $\Delta \tau_{\rm RMS}$ across all regions, weighted by the number of stars with age measurements in each region.

The best (lowest-scoring) model in the age--[O/Fe] plane is the triple-system DTD with the two-infall SFH. Models which score almost as well are the plateau ($W=1$ Gyr) and triple-system DTDs with either the early-burst or two-infall SFH. Visually, the non-monotonic bump in the age--[O/Fe] relation produced by the late-burst SFH does not match the observed distribution, but it actually {\it improves} $\Delta \tau_{\rm RMS}$ by lowering the median age of stars in the low-[O/Fe] bins. If the shape of the distribution is taken into account, the late-burst SFH produces the worst match to the data.
We discuss the quantitative scores in further Section \ref{sec:discussion-scores} below.

\section{Discussion}
\label{sec:discussion}

\subsection{Qualitative Comparisons}
\label{sec:discussion-scores}

\begin{table*}
\centering
\caption{Qualitative summary of comparisons between the model output,
APOGEE DR17 abundances, and \citetalias{Leung2023-Ages} ages for each multi-zone
model. See discussion in Section \ref{sec:feh-df} for the [Fe/H] DF, 
Section \ref{sec:ofe-df} for the [O/Fe] DF, Section \ref{sec:bimodality} for
the [O/Fe] bimodality, Section \ref{sec:ofe-feh} for the [O/Fe]--[Fe/H] plane,
and Section \ref{sec:age-ofe} for the age--[O/Fe] plane. Table \ref{tab:scores}
presents the quantitative scores used to make these comparisons.}
\label{tab:results}
  \begin{tabular}{ll|ccccc}
\hline\hline
DTD & SFH & MDF & [O/Fe] DF & [O/Fe] Bimodality & [Fe/H]--[O/Fe] & Age--[O/Fe] \\
\hline
Two-population & Inside-out & \no & \no & \no & \no & \no \\
($t_p=0.05$ Gyr) & Late-burst & \no & \no & \no & \no & \no \\
 & Early-burst & \yes & \no & \yes & \no & \meh \\
 & Two-infall & \meh & \meh & \no & \no & \no \\ 
\hline
Power-law & Inside-out & \no & \no & \no & \no & \no \\
($\alpha=-1.4$) & Late-burst & \no & \no & \no & \no & \no \\
 & Early-burst & \meh & \no & \yes & \no & \no \\
 & Two-infall & \no & \no & \no & \no & \no \\ 
\hline
Power-law & Inside-out & \no & \meh & \no & \meh & \no \\
($\alpha=-1.1$) & Late-burst & \no & \no & \yes & \meh & \no \\
 & Early-burst & \meh & \no & \yes & \no & \meh \\
 & Two-infall & \yes & \meh & \yes & \meh & \no \\ 
\hline
Exponential & Inside-out & \meh & \meh & \no & \meh & \no \\
($\tau=1.5$ Gyr) & Late-burst & \meh & \yes & \yes & \meh & \meh \\
 & Early-burst & \no & \no & \yes & \no & \meh \\
 & Two-infall & \yes & \meh & \yes & \meh & \meh \\ 
\hline
Exponential & Inside-out & \yes & \yes & \no & \yes & \meh \\
($\tau=3.0$ Gyr) & Late-burst & \meh & \yes & \yes & \yes & \yes \\
 & Early-burst & \meh & \no & \yes & \meh & \yes \\
 & Two-infall & \yes & \meh & \yes & \meh & \yes \\ 
\hline
Plateau & Inside-out & \meh & \yes & \no & \yes & \meh \\
($W=0.3$ Gyr) & Late-burst & \no & \yes & \yes & \yes & \meh \\
 & Early-burst & \meh & \meh & \yes & \no & \yes \\
 & Two-infall & \yes & \yes & \yes & \yes & \meh \\ 
\hline
Plateau & Inside-out & \yes & \yes & \yes & \yes & \meh \\
($W=1.0$ Gyr) & Late-burst & \no & \yes & \yes & \yes & \yes \\
 & Early-burst & \yes & \meh & \yes & \meh & \yes \\
 & Two-infall & \yes & \yes & \yes & \yes & \yes \\ 
\hline
Triple-system & Inside-out & \yes & \yes & \no & \yes & \yes \\
($t_{\rm rise}=0.5$ Gyr) & Late-burst & \no & \yes & \yes & \yes & \yes \\
 & Early-burst & \meh & \meh & \yes & \meh & \yes \\
 & Two-infall & \yes & \meh & \yes & \yes & \yes \\
\hline
\end{tabular}
\unskip\label{output/summary_table.tex}\unskip%

\end{table*}

In Section \ref{sec:multizone-results}, we focused on a representative subset of our 32 multi-zone models (four SFHs and eight DTDs). Here, we compare all of our model outputs to APOGEE across five observables: the MDF, [O/Fe] DF, [O/Fe]--[Fe/H] plane, age--[O/Fe] plane, and [O/Fe] bimodality. We perform statistical tests between APOGEE and the model outputs in each region of the Galaxy as described in corresponding subsections of Section \ref{sec:multizone-results}, then compute the average weighted by the size of the APOGEE sample in each region to obtain a single numerical score.

The relative performance of each model is summarized in Table \ref{tab:results}. We use these scores to indicate combinations of SFH and DTD that are favorable or unfavorable in certain regimes, but we do {\it not} fit our models to the data due to computational expense. To avoid drawing strong conclusions from small numerical differences in scores, we simply write \yes, \meh, or \no, which corresponds to a score in the top, middle, or bottom third out of all models, respectively. The exact numerical scores are presented in Appendix \ref{app:quantitative-scores}.

Some of the variation between models can be explained by the choice of SFH. The two-infall models tend to out-perform the others for the MDFs, while the late-burst models score poorly, especially with the prompt DTDs. The early-burst models consistently have the lowest scores for the [O/Fe] DF and [O/Fe]--[Fe/H] distribution, but are able to produce a bimodal [O/Fe] distribution with every DTD (see discussion in Section \ref{sec:bimodality}). The late-burst and two-infall SFHs also produce a bimodal [O/Fe] distribution with all DTDs except those with the highest prompt fraction, while the inside-out models never produce bimodality. The inside-out models also tend score poorly in the age--[O/Fe] plane, while the early-burst models tend to score well, although as discussed in Section \ref{sec:age-ofe}, adjusting the time of the peak SFR or running the models for a shorter period of time would affect the level of agreement in the high-[O/Fe] bins.

It is somewhat surprising that the early-burst models score poorly against the APOGEE [O/Fe] DFs, given that they produce the clearest bimodal distributions. The KL divergence test heavily penalizes models with a high density in a region where the observations have little, as is the case for the high-$\alpha$ sequence in the outer Galaxy and close to the midplane (see Figure \ref{fig:ofe-df-sfh}). This similarly explains the early-burst models' poor performance in the [O/Fe]--[Fe/H] plane. An iteration of this SFH where the early burst predominantly affects the inner galaxy is probably more accurate and might have more success at reproducing the [O/Fe] DF across the disk.

The choice of DTD has a clear effect on the model scores, and this effect is similar for most of the observables. The models which perform the best (most \yes's and fewest \no's) are the most extended DTDs with the fewest prompt SNe Ia: both plateau DTDs, the exponential DTD with $\tau=3$ Gyr, and the triple-system DTD. The latter actually produces the highest scores for each observable, but the plateau DTD with $W=1$ Gyr is the most successful across all SFHs; both models have some of the longest median delay times. Models with a large fraction of prompt SNe Ia, such as the power-law and two-population DTDs, fare quite poorly, with the steepest power-law ($\alpha=-1.4$) and two-population DTDs ending up in the bottom third across the board for most of our SFHs. The fiducial power-law ($\alpha=-1.1$) does slightly better, but still compares poorly to the more extended DTDs.

Each DTD tends to score similarly across the board, but there are some combinations of SFH and DTD that buck the general trend. For example, the two-population DTD with the early-burst SFH produces an MDF which scores relatively well. The early-burst models generally produce MDFs in the \meh category, so a small increase in the numerical score bumps it up to \yes; this indicates the insensitivity of the MDF to the DTD in general. The exponential DTD with $\tau=1.5$ Gyr has generally middling performance, but does a notably poorer job when combined with the early-burst SFH, a result of the generally poor performance of that SFH. 

The plateau DTD with $W=1$ Gyr, our most successful model overall, poorly reproduces the MDF with the late-burst SFH, while the exponential DTD with $\tau=3$ Gyr produces better agreement with the data for that SFH. Finally, the inside-out SFH generally does not reproduce the APOGEE age--[O/Fe] relation well, but it scores better than average when combined with the triple-system DTD.

Our model scores are highly sensitive to small changes in the nucleosynthetic yields. A decrease in the SN Ia yield of Fe to $y_{\rm Fe}^{\rm Ia}=0.0017$, which shifts the end-point of the gas abundance tracks up by $\sim+0.05$ dex in [O/Fe], produces dramatically different scores for many of the models. This is because the KL divergence tests penalize distributions which are not well aligned with the data, even if the general trends and shape of the distribution are reproduced. For example, if the two-infall models are run with $y_{\rm Fe}^{\rm Ia}=0.0017$, the abundance tracks do not dip below solar [O/Fe] (see the bottom-right panel of Figure \ref{fig:ofe-feh-sfh}) and consequently they out-score every other SFH. Small adjustments in the yields can affect the quality of the fit between our models and the data, so we caution against over-interpreting the qualitative comparisons in Table \ref{tab:results}.

We also run a two-sample Kolmogorov-Smirnov (KS) test on the model [Fe/H] and [O/Fe] DFs to estimate the significance of the agreement with the observed distributions. However, we consistently reject the null hypothesis that the model and observed abundances are drawn from the same distribution at very high significance ($p<<0.05$) in each Galactic region. The large sample size means that even small deviations from the APOGEE distribution result in a very small $p$-value, making the KS test a poor diagnostic for model comparison. Overall, this reinforces the conclusion that even our best-performing models cannot reproduce all observations.

\subsection{The Two-Infall SFH}
\label{sec:two-infall-discussion}

There have been many comparative GCE studies of the DTD with the two-infall model, providing an important point of comparison with our models. For example, \citet{Matteucci2006-BimodalDTDConsequences} explored the consequences of the two-population DTD \citep{Mannucci2006-TwoPopulations}, finding that its very high prompt SN Ia rate began to pollute the ISM during the halo phase and led to a faster decline in [O/Fe] with [Fe/H]. \citet{Matteucci2009-DTDModels} compared several DTDs, including the analytic forms of \citet{Greggio2005-AnalyticalRates} and the two-population DTD, in a multi-zone GCE model of the disk. Their comparisons to data were limited to the solar neighborhood, and unlike our models, they did not factor in radial migration or gas outflows. Nevertheless, their conclusions align fairly well with ours: a relatively low fraction of prompt SNe Ia is needed to produce good agreement with observations.

More recently, \citet{Palicio2023-AnalyticDTD} compared a similar suite of DTDs in one-zone models with a two-infall SFH. In contrast to previous studies of the two-infall model \citep[e.g.,][]{Chiappini1997-TwoInfall,Matteucci2009-DTDModels,Spitoni2021-TwoInfall}, they did incorporate gas outflows, making their models especially well-suited to compare to ours. By modifying their yields, outflow mass-loading factor, and some of the parameters of their SFH, \citet{Palicio2023-AnalyticDTD} were able to achieve a good fit to solar neighborhood abundance data for both the SD and DD analytic DTDs, which are approximated by our exponential ($\tau=1.5$ Gyr) and plateau ($W=1$ Gyr) models, respectively. Our results and theirs highlight the need for independent constraints on the SFH to resolve degeneracies with the DTD.

To our knowledge, this paper is the first exploration of the two-infall SFH in a multi-zone GCE model which incorporates both mass-loaded outflows and radial migration. A detailed examination of the parameters of the two-infall model is beyond the scope of this paper but will be the subject of future work.

\subsection{Extragalactic Constraints}

The power-law ($\alpha=-1.1$) DTD has the strongest observational motivation but poorly reproduces the disk abundance distributions. This can be mitigated somewhat with a longer minimum delay time, which has a similar effect on chemical evolution tracks as the addition of an initial plateau in the DTD (see discussion in Section \ref{sec:onezone-results}). Even so, it is clear that the high fraction of prompt SNe Ia in extragalactic constraints on the DTD by, e.g., \citet{Maoz2017-CosmicDTD} is at odds with Galactic chemical abundance measurements. 

This tension could suggest that the Milky Way obeys a different DTD than other galaxies. This would not be too far beyond \citeauthor{Maoz2017-CosmicDTD}'s \citeyearpar{Maoz2017-CosmicDTD} finding that field galaxies and galaxy clusters have a different DTD slope. However, \citet{Walcher2016-SelfSimilarity} argued that the similarity of the age--\aFe relation between solar neighborhood stars and nearby elliptical galaxies is evidence for a universal DTD. A physical mechanism would be needed to produce a different slope or form for the DTD in different environments, such as a metallicity dependence in the fraction of close binaries \citep[e.g.,][]{Moe2019-CloseBinaryFraction}.

On the other hand, the difference between constraints from GCE models and extragalactic surveys indicates that these types of studies are most sensitive to different regimes of the DTD. Our results demonstrate that the high-$\alpha$ sequence in GCE models is highly sensitive to the DTD at short delay times. Measurements of galactic or cosmic SFHs typically provide constraints for the DTD in coarse age bins, with especially large uncertainties in the youngest bins \citep[e.g.,][]{MaozMannucci2012-SNeIaReview}, and it is difficult to constrain the SFH of individual galaxies at long lookback times \citep{Conroy2013-PanchromaticSED}. Additionally, measurements of the cosmic SN Ia rate become considerably uncertain at $z\gtrsim1$ \citep[see, e.g.,][]{Palicio2024-CosmicSNIaRate}. As a result, constraints from external galaxies should be more sensitive to the DTD at long delay times.

\citet{Palicio2024-CosmicSNIaRate} fit combinations of cosmic star formation rates (CSFRs) and DTDs, many of which are similar to the forms in this paper, to the observed cosmic SN Ia rate. Notably, the DTD that best fit the majority of their CSFRs was the single-degenerate DTD of \citet{MatteucciRecchi2001-SNIaTimescale}, which is similar to the exponential form with $\tau=1.5$ Gyr (see Appendix \ref{app:analytic-dtds} for more discussion). They were able to exclude DTDs with a very high or very low fraction of prompt SNe Ia, but a number of their DTDs could produce a convincing fit to the observed rates with the right CSFR. Despite a very different methodology, their results mirror ours: that many forms for the DTD can produce a reasonable fit to the data when combined with the right SFH. 

\section{Conclusions}
\label{sec:conclusions}

We have explored the consequences of eight different forms for the SN Ia DTD in multi-zone GCE models with radial migration. For each DTD, we explored combinations with four different popular SFHs from the literature, which represent a broad range of behavior over the lifetime of the disk seen in many prior GCE models. We compared our model outputs to abundances from APOGEE and ages from \citetalias{Leung2023-Ages} for stars across the Milky Way disk. For each model, we computed a numerical score that reflects the agreement between the predictions and data across the entire disk for five observables. Our main conclusions are as follows:

\begin{itemize}
    \item While some combinations of SFH and DTD perform better than others, none of our models are able to reproduce every observed feature of the Milky Way disk.
    
    \item The plateau DTD with a width $W=1$ Gyr is best able to reproduce the observed abundance patterns for three of the four SFHs. For the inside-out SFH, it is narrowly surpassed by the (similar) triple-system DTD.

    \item In general, we favor a DTD with a small fraction of prompt SNe Ia. The models with exponential, plateau, and triple-system DTDs perform significantly better than the models with two-population and power-law DTDs across all four SFHs.

    \item The observationally-derived $t^{-1.1}$ power-law DTD produces too few high-$\alpha$ stars. This could be mitigated with a longer minimum delay time or the addition of an initial plateau in the DTD at short delay times.
    
    \item The SFH is the critical factor for producing a bimodal \aFe distribution at fixed [Fe/H]. On its own, the DTD cannot produce a bimodal \aFe distribution that matches what is observed. However, it does affect the location and strength of the high-$\alpha$ sequence, potentially enhancing the \aFe bimodality resulting from the choice of SFH.
\end{itemize}

The origin of the Milky Way's \aFe bimodality remains disputed. Some authors have argued that the combination of inside-out growth and radial migration sufficiently explains the observed distribution \citep[e.g.,][]{Kubryk2015-RadialMigrationEvolution,Sharma2021-RadialMigration,Chen2023-RadialMixingRedux,Prantzos2023-ThinThickDisks}, while others have argued that multiple episodes of gas infall at early times are required \citep[e.g.,][]{Chiappini1997-TwoInfall,Mackereth2018-AlphaBimodality,Spitoni2019-TwoInfall,Spitoni2020-TwoInfall,Spitoni2021-TwoInfall}. In agreement with \citetalias{Johnson2021-Migration}, we find that a smooth SFH combined with radial migration does not suffice. We find that these parameter choices predict too many stars between the high- and low-$\alpha$ sequences, resulting in a broad but unimodal \aFe distribution at fixed [Fe/H].

We found that the MDF is least able to provide constraints on the DTD. The MDF is more sensitive to the SFH, but overall trends across the Galaxy are primarily driven by the assumed radial abundance gradient and stellar migration prescription. However, the MDF is more sensitive to the DTD in the inner Galaxy due to the more sharply declining SFH (see discussion in Section \ref{sec:feh-df}). Under the model of inside-out formation, the MDF in the inner Galaxy traces older populations which are more sensitive to the enrichment of prompt SNe Ia.

We implemented a stellar migration scheme which reproduces the abundance trends seen in the models of \citetalias{Johnson2021-Migration}, but produces smoother abundance distributions. Our method is flexible and is not tied to the output of a single hydrodynamical simulation. In future work, we will explore the effect of the strength and speed of radial migration on GCE models.

Recent studies have shown that the high specific SN Ia rates observed in low-mass galaxies \citep[e.g.,][]{Brown2019-ASASSNrates,Wiseman2021-DESRates} can be explained by a metallicity-dependent rate of SNe Ia \citep{Gandhi2022-MetallicityDependentRates,Johnson2023-Binaries}. A similar metallicity dependence has also been observed in the rate of CCSNe \citep{Pessi2023-MetalDepCCSNe}. These previous investigations varied only the normalization in the DTD. \citet{Gandhi2022-MetallicityDependentRates} take into account radial migration by construction through their use of the FIRE-2 simulations. An exploration in the context of multi-zone models would be an interesting direction for future work, as would variations in the DTD shape.

Our results indicate that the allowed range of parameter space in GCE models is still too broad to precisely constrain the DTD. Future constraints may come from the Legacy Survey of Space and Time (LSST) at the Vera Rubin Observatory \citep{Ivezic2019-LSST}, which is expected to observe several million SNe during its 10-year run. On the other hand,
the improved sample size of SDSS-V \citep{Kollmeier2017-SDSS-V} will enable future GCE studies to constrain both the Galactic SFH and the DTD at a higher confidence.

\section*{Acknowledgements}

We thank Prof.\ David H.\ Weinberg and attendees of OSU's Galaxy Hour for many useful discussions over the course of this project. We also thank the anonymous reviewer for their thoughtful and constructive comments on the manuscript.

LOD and JAJ acknowledge support from National Science Foundation grant no.\ AST-2307621. JAJ and JWJ acknowledge support from National Science Foundation grant no.\ AST-1909841.
LOD acknowledges financial support from an Ohio State University Fellowship.
JWJ acknowledges financial support from an Ohio State University Presidential Fellowship and a Carnegie Theoretical Astrophysics Center postdoctoral fellowship.

Funding for the Sloan Digital Sky 
Survey IV has been provided by the 
Alfred P.\ Sloan Foundation, the U.S.\ 
Department of Energy Office of 
Science, and the Participating 
Institutions. 

SDSS-IV acknowledges support and 
resources from the Center for High 
Performance Computing  at the 
University of Utah. The SDSS 
website is \url{www.sdss4.org}.

SDSS-IV is managed by the 
Astrophysical Research Consortium 
for the Participating Institutions 
of the SDSS Collaboration including 
the Brazilian Participation Group, 
the Carnegie Institution for Science, 
Carnegie Mellon University, Center for 
Astrophysics | Harvard \& 
Smithsonian, the Chilean Participation 
Group, the French Participation Group, 
Instituto de Astrof\'isica de 
Canarias, The Johns Hopkins 
University, Kavli Institute for the 
Physics and Mathematics of the 
Universe (IPMU) / University of 
Tokyo, the Korean Participation Group, 
Lawrence Berkeley National Laboratory, 
Leibniz Institut f\"ur Astrophysik 
Potsdam (AIP),  Max-Planck-Institut 
f\"ur Astronomie (MPIA Heidelberg), 
Max-Planck-Institut f\"ur 
Astrophysik (MPA Garching), 
Max-Planck-Institut f\"ur 
Extraterrestrische Physik (MPE), 
National Astronomical Observatories of 
China, New Mexico State University, 
New York University, University of 
Notre Dame, Observat\'ario 
Nacional / MCTI, The Ohio State 
University, Pennsylvania State 
University, Shanghai 
Astronomical Observatory, United 
Kingdom Participation Group, 
Universidad Nacional Aut\'onoma 
de M\'exico, University of Arizona, 
University of Colorado Boulder, 
University of Oxford, University of 
Portsmouth, University of Utah, 
University of Virginia, University 
of Washington, University of 
Wisconsin, Vanderbilt University, 
and Yale University.

This work has made use of data from the European Space Agency (ESA) mission
{\it Gaia} (\url{https://www.cosmos.esa.int/gaia}), processed by the {\it Gaia}
Data Processing and Analysis Consortium (DPAC,
\url{https://www.cosmos.esa.int/web/gaia/dpac/consortium}). Funding for the DPAC
has been provided by national institutions, in particular the institutions
participating in the {\it Gaia} Multilateral Agreement.

We would like to acknowledge the land that The Ohio State University occupies is the ancestral and contemporary territory of the Shawnee, Potawatomi, Delaware, Miami, Peoria, Seneca, Wyandotte, Ojibwe and many other Indigenous peoples. Specifically, the university resides on land ceded in the 1795 Treaty of Greeneville and the forced removal of tribes through the Indian Removal Act of 1830. As a land grant institution, we want to honor the resiliency of these tribal nations and recognize the historical contexts that has and continues to affect the Indigenous peoples of this land.

\software{\vice \citep{JohnsonWeinberg2020-Starbursts}, Astropy \citep{astropy2013,astropy2018,astropy2022}, scikit-learn \citep{Pedregosa2011-ScikitLearn}, SciPy \citep{2020SciPy-NMeth}, Matplotlib \citep{Hunter2007-Matplotlib}}

\appendix

\section{Reproducibility}
\label{app:reproducibility}

This study was carried out using the reproducibility software
\href{https://github.com/showyourwork/showyourwork}{\showyourwork}
\citep{Luger2021-showyourwork}, which leverages continuous integration to
programmatically download the data from
\href{https://zenodo.org/}{zenodo.org}, create the figures, and
compile the manuscript. Each figure caption contains two links: one
to the dataset stored on zenodo used in the corresponding figure,
and the other to the script used to make the figure (at the commit
corresponding to the current build of the manuscript). The git
repository associated to this study is publicly available at
\url{\GitHubURL}, and the release v1.1.1 allows anyone to re-build the entire 
manuscript. The multi-zone model outputs and APOGEE sample dataset are stored at \url{https://zenodo.org/doi/10.5281/zenodo.10961090}, and the source code for this manuscript is stored at \url{https://zenodo.org/doi/10.5281/zenodo.12521399}.

\section{Analytic DTDs}
\label{app:analytic-dtds}

\begin{figure*}
    \centering
    \includegraphics[width=0.49\linewidth]{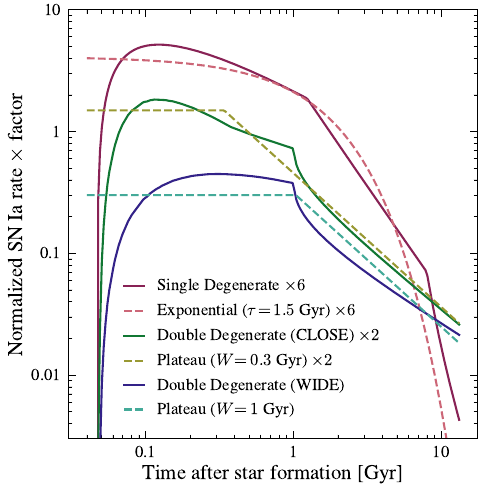}
    \includegraphics[width=0.49\linewidth]{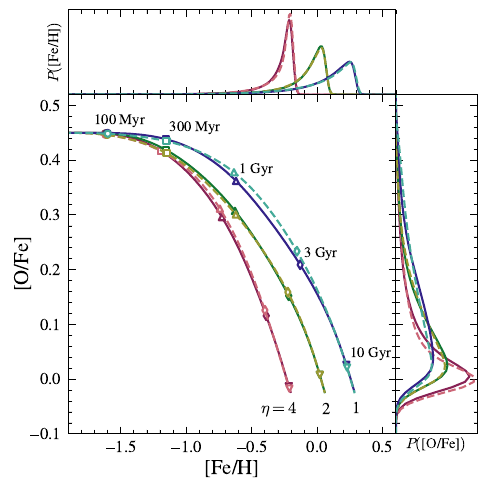}
    \caption{\textit{Left:} Analytic DTDs from \citet[][solid curves]{Greggio2005-AnalyticalRates} and simplified approximations thereof (dashed curves; see Section \ref{sec:dtd-models}). Some functions are presented with a constant multiplicative factor for visual clarity. \textit{Right:} Abundance tracks and distributions from one-zone models with the analytic and simple DTDs (same color scheme). For visual clarity, we vary the mass-loading factor to be $\eta=4$, $\eta=2$, and $\eta=1$ for the red, green, and blue curves, respectively. All other model parameters are identical.}
    \label{fig:analytic-dtds}
    \script{analytical_dtd_twopanel.py}
\end{figure*}

\citet{Greggio2005-AnalyticalRates} derived analytic DTDs for SD and DD progenitor systems from assumptions about binary stellar evolution and mass exchange. Significant parameters for the shape of the DTD are the distribution and range of stellar masses in progenitor systems, the efficiency of accretion in the SD scenario, and the distribution of separations at birth in the DD scenario. The left-hand panel of Figure \ref{fig:analytic-dtds} shows several of the \citet{Greggio2005-AnalyticalRates} analytic DTDs: one for SD progenitors, and two different prescriptions for DD progenitors (``WIDE'' and ``CLOSE''). The difference between the DD prescriptions relates to the ratio between the separation of the DD system and the initial separation of the binary, $A/A_0$. In the ``WIDE'' scheme, it is assumed that $A/A_0$ spans a wide distribution, and that the distributions of $A$ and total mass of the system $m_{\rm DD}$ are independent, so one cannot necessarily predict the total merge time of a system based on its initial parameters. In the ``CLOSE'' scheme, there is assumed to be a narrow distribution of $A/A_0$ and a correlation between $A$ and $m_{\rm DD}$, so the most massive binaries tend to merge quickly and the least massive merge last.

Here we state our assumptions for the hyper-parameters which can affect the shape of the \citet{Greggio2005-AnalyticalRates} DTDs. For the SD case, we assume a power law slope of the mass ratio distribution with $\gamma=1$, an efficiency of mass transfer $\epsilon=1$, and maximum initial primary mass of $8~{\rm M}_\odot$. For the DD channel, we additionally assume a nuclear timescale for the least massive secondary $\tau_{\rm n,x}=1$ Gyr, a minimum gravitational insprial delay $\tau_{\rm gw,i}=1$ Myr, an exponent of the power law distribution of final separations $\beta_{\rm a}=0$ (for the WIDE scheme), and an exponent of the power law distribution of gravitational delays $\beta_{\rm g}=-0.75$ (for the CLOSE scheme).

In the left-hand panel of Figure \ref{fig:analytic-dtds}, we also include simple functions which approximate the analytic DTDs of \citet{Greggio2005-AnalyticalRates}. Chemical abundance distributions are sensitive to the broad shape of the DTD but are agnostic to the underlying physics of the progenitor systems. These simplified forms reduce the number of free parameters for the DTD and make the GCE model predictions easier to interpret.

The right-hand panel of Figure \ref{fig:analytic-dtds} shows the results of one-zone chemical evolution models with the \citet{Greggio2005-AnalyticalRates} DTDs and our simplified forms. We use the same model parameters as in Section \ref{sec:onezone-results} but with different values of $\eta$ to spread the tracks out visually in [Fe/H]. The model with the SD DTD follows a nearly identical track to the exponential ($\tau=1.5$ Gyr) DTD, and they produce very similar distributions of [O/Fe]. Likewise, the DD CLOSE DTD is well approximated by the plateau DTD with $W=0.3$ Gyr and a power-law slope $\alpha=-1.1$. The WIDE prescription is also best approximated by a plateau DTD, but with a longer plateau width of $W=1$ Gyr. In all cases, the effect of the difference between the analytic DTD and its simple approximation is too small to be observed. We also ran a multi-zone model with the inside-out SFH and the \citet{Greggio2005-AnalyticalRates} SD DTD and found it produced nearly identical results to the model with the exponential ($\tau=1.5$ Gyr) DTD.

\section{Stellar Migration}
\label{app:migration}

In their multi-zone models, \citetalias{Johnson2021-Migration} randomly assign an analogue star particle from \hydro, adopting its radial migration distance $\Delta R$ and final midplane distance $z$, for each stellar population generated by \vice. The analogues are chosen such that the star particle was born at a similar radius and time as the stellar population in the GCE model. This prescription allows \vice to adopt a realistic pattern of radial migration without needing to implement its own hydrodynamical simulation. However, in regions where the number of \hydro star particles is relatively low, such as at large $R_{\rm gal}$ and small $t$, a single \hydro star particle can be assigned as an analogue to multiple \vice stellar populations. These populations will have similar formation and migration histories and consequently similar abundances, which produces unphysical ``clumps'' of stars in the abundance distributions at large $|z|$ and $R_{\rm gal}$.

We adopt a prescription for radial migration which produces smoother abundance distributions while still following the behavior of \hydro. We fit a Gaussian to the distribution of $\Delta R = R_{\rm final} - R_{\rm form}$ from the \hydro output, binned by both formation radius $R_{\rm form}$ and age. We are motivated by the findings of \citet{Okalidis2022-AurigaMigration} that the strength of stellar migration in the Auriga simulations \citep{Grand2017-AurigaSims} varies with both $R_{\rm form}$ and age. Each Gaussian is centered at 0, and we find that the scale $\sigma_{\rm RM}$ is best described by the function
\begin{equation}
    \sigma_{\rm RM} = \sigma_{\rm RM8} \Big(\frac{\tau}{8\,{\rm Gyr}}\Big)^{0.33} \Big(\frac{R_{\rm form}}{8\,{\rm kpc}}\Big)^{0.61}
    \label{eq:radial-migration}
\end{equation}
where $\tau$ is the stellar age and $\sigma_{\rm RM8}=2.68$ kpc describes the migration strength for an 8 Gyr old population with $R_{\rm form}=8$ kpc. 
For comparison, \citet{Frankel2018-RadialMigration} found a steeper $\tau$-dependence of $\sigma_{\rm RM}\propto \tau^{1/2}$ and a higher scaling of $\sigma_{\rm RM8}=3.6$ kpc for a sample of APOGEE red clump stars.
Our age scaling is in good agreement with \citet{Lu2023-LMCStars}, who find that radial migration in galaxies from the NIHAO simulations \citep{Wang2015-NIHAOSims} follow a relatively universal relation of $\sigma_{\rm RM}\propto\tau^{0.32}$, but with a slightly higher $\sigma_{\rm RM8}\approx3$ kpc. We use the lower value here as it reproduces the trends in \hydro, and by extension \citetalias{Johnson2021-Migration}, but the $\sim25\%$ difference may affect the predictions.

When \vice forms a stellar population at initial radius $R_{\rm form}$, we assign a value of $\Delta R$ by randomly sampling from a Gaussian with a width given by Equation \ref{eq:radial-migration}. The star particle migrates to its final radius $R_{\rm final}$ in a similar manner to the ``diffusion'' case from \citetalias{Johnson2021-Migration}, but with a time dependence $\propto \Delta t^{1/3}$, motivated by the age-scaling of $\sigma_{\rm RM}$.

We note that the \hydro galaxy has a weak and transient bar, in contrast to the Milky Way. The presence of a strong bar has been found to affect the strength of radial migration throughout the disk \citep[e.g.,][]{Brunetti2011-BarredSpiralDiffusion} and lead to a flattening of the metallicity gradient for old populations \citep{Okalidis2022-AurigaMigration}.

\begin{figure*}
    \centering
    \includegraphics[width=\linewidth]{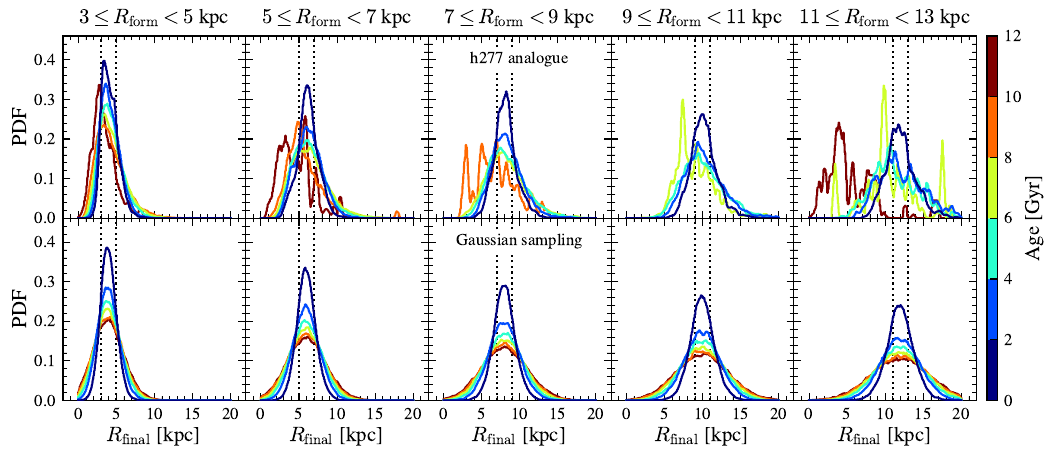}
    \caption{The distribution of final radius $R_{\rm final}$ as a function of formation radius $R_{\rm form}$ and age for the \hydro analogue (top row) and Gaussian sampling scheme (bottom row; see discussion in Appendix \ref{app:migration}). From left to right, star particles are binned by formation annulus, as noted at the top of each column of panels. Within each panel, colored curves represent the different age bins, ranging from the youngest stars (dark blue) to the oldest (dark red). In the top row, we exclude age bins with fewer than 100 unique analogue IDs for visual clarity. All distributions are normalized so that the area under the curve is 1, and have been boxcar-smoothed with a window width of 0.5 kpc. The vertical dotted black lines indicate the bounds of each bin in $R_{\rm form}$; stars within that region of the distribution have not migrated significantly far from their birth radius.}
    \label{fig:radial-migration}
    \script{radial_migration.py}
\end{figure*}

\begin{figure*}
    \centering
    \includegraphics[width=\linewidth]{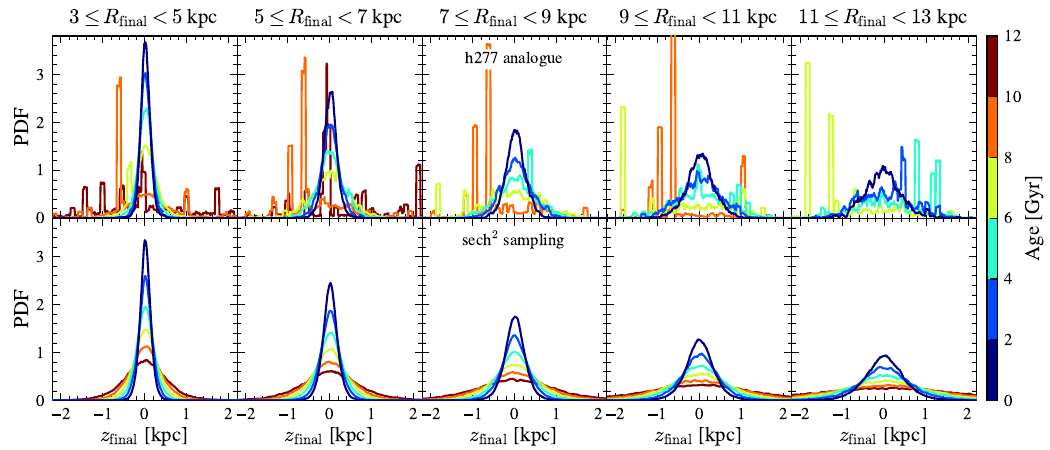}
    \caption{Similar to Figure \ref{fig:radial-migration} but for the distribution of present-day midplane distance $z_{\rm final}$ as a function of radius and age. From left to right, star particles are binned by \textit{final} annulus. In the top row, we exclude age bins with fewer than 500 unique analogue IDs for clarity. All distributions have been boxcar-smoothed with a window width of 0.1 kpc.}
    \label{fig:midplane-distance}
    \script{midplane_distance.py}
\end{figure*}

Figure \ref{fig:radial-migration} compares the distributions of $R_{\rm final}$ in bins of $R_{\rm form}$ and stellar age between the \hydro analogue method and our new prescription. There is good agreement across the Galaxy in the youngest age bins, but the ``clumpiness'' of the \hydro analogue populations, a consequence of sampling noise, becomes evident for old stars formed in the outer Galaxy. The distribution of \hydro star particles in the $10\leq\tau<12$ Gyr and $11\leq R_{\rm form}<13$ kpc bin indicates significant inward migration due to a merging satellite. Our Gaussian sampling scheme eliminates both the clumpiness and the impact of mergers and other external events on radial migration.

Like \citetalias{Johnson2021-Migration}, we assume all stellar populations form in the midplane ($z=0$). \citetalias{Johnson2021-Migration} take the present-day midplane distance $z_{\rm final}$ directly from the \hydro analogue particle. To produce smoother abundance distributions, we fit a sech$^2$ function \citep{Spitzer1942} to the distribution of $z$ in \hydro. The PDF of $z_{\rm final}$ given some scale height $h_z$ is
\begin{equation}
    {\rm PDF}(z_{\rm final}) = \frac{1}{4 h_z} {\rm sech}^2\Big(\frac{z_{\rm final}}{2 h_z}\Big).
    \label{eq:sech-pdf}
\end{equation}
We fit Equation \ref{eq:sech-pdf} to the distributions of $z$ in \hydro in varying bins of $T$ and $R_{\rm final}$. We find that $h_z$ is best described by the function
\begin{equation}
    h_z = (h_{z,s}/e^2) \exp(\tau/\tau_s + R_{\rm final}/R_s)
    \label{eq:scale-height}
\end{equation}
where $h_{z,s}=0.24$ kpc is the scale height at $\tau_s=7$ Gyr and $R_s=6$ kpc.
For each star particle in \vice, we sample $z_{\rm final}$ from the distribution described by Equation \ref{eq:sech-pdf} with a width given by Equation \ref{eq:scale-height}. Figure \ref{fig:midplane-distance} shows the resulting distributions of $z_{\rm final}$ are similar to the \hydro analogue scheme for all but the oldest stellar populations.

\section{Quantitative Comparison Scores}
\label{app:quantitative-scores}

\begin{table*}
\centering
\caption{Quantitative scores comparing the model output,
APOGEE DR17 abundances, and \citetalias{Leung2023-Ages} ages for each multi-zone
model. See discussion in Section \ref{sec:feh-df} for the [Fe/H] DF, 
Section \ref{sec:ofe-df} for the [O/Fe] DF, Section \ref{sec:ofe-feh} for the 
[O/Fe]--[Fe/H] plane, and Section \ref{sec:age-ofe} for the age--[O/Fe] 
plane. Table \ref{tab:results} summarizes the relative performance
of each model based on the scores presented here.}
\label{tab:scores}
  \begin{tabular}{ll|cccc}
\hline\hline
DTD & SFH & MDF & [O/Fe] DF & [Fe/H]--[O/Fe] & Age--[O/Fe] \\
\hline
Two-population & Inside-out & 0.292 & 0.947 & 1.992 & 3.41 \\
($t_p=0.05$ Gyr) & Late-burst & 0.291 & 0.99 & 1.996 & 3.069 \\
 & Early-burst & 0.211 & 1.591 & 2.6 & 2.439 \\
 & Two-infall & 0.219 & 0.638 & 1.839 & 2.873 \\ 
\hline
Power-law & Inside-out & 0.343 & 1.318 & 2.695 & 4.211 \\
($\alpha=-1.4$) & Late-burst & 0.327 & 1.35 & 2.73 & 3.608 \\
 & Early-burst & 0.233 & 1.73 & 3.424 & 3.084 \\
 & Two-infall & 0.262 & 0.898 & 2.46 & 3.498 \\ 
\hline
Power-law & Inside-out & 0.288 & 0.688 & 1.513 & 3.255 \\
($\alpha=-1.1$) & Late-burst & 0.292 & 0.731 & 1.552 & 2.956 \\
 & Early-burst & 0.215 & 1.197 & 2.471 & 2.257 \\
 & Two-infall & 0.202 & 0.437 & 1.327 & 2.56 \\ 
\hline
Exponential & Inside-out & 0.24 & 0.588 & 1.119 & 3.158 \\
($\tau=1.5$ Gyr) & Late-burst & 0.222 & 0.388 & 0.921 & 2.54 \\
 & Early-burst & 0.247 & 0.954 & 2.027 & 2.224 \\
 & Two-infall & 0.18 & 0.482 & 1.156 & 2.283 \\ 
\hline
Exponential & Inside-out & 0.183 & 0.328 & 0.773 & 1.976 \\
($\tau=3.0$ Gyr) & Late-burst & 0.214 & 0.253 & 0.682 & 1.788 \\
 & Early-burst & 0.217 & 0.732 & 1.391 & 1.52 \\
 & Two-infall & 0.134 & 0.484 & 0.897 & 1.518 \\ 
\hline
Plateau & Inside-out & 0.235 & 0.353 & 0.837 & 2.509 \\
($W=0.3$ Gyr) & Late-burst & 0.256 & 0.372 & 0.854 & 2.241 \\
 & Early-burst & 0.215 & 0.663 & 1.656 & 1.731 \\
 & Two-infall & 0.154 & 0.349 & 0.863 & 1.845 \\ 
\hline
Plateau & Inside-out & 0.199 & 0.243 & 0.679 & 1.844 \\
($W=1.0$ Gyr) & Late-burst & 0.243 & 0.314 & 0.73 & 1.798 \\
 & Early-burst & 0.211 & 0.567 & 1.207 & 1.399 \\
 & Two-infall & 0.126 & 0.422 & 0.799 & 1.37 \\ 
\hline
Triple-system & Inside-out & 0.179 & 0.231 & 0.678 & 1.568 \\
($t_{\rm rise}=0.5$ Gyr) & Late-burst & 0.248 & 0.327 & 0.758 & 1.569 \\
 & Early-burst & 0.223 & 0.603 & 1.239 & 1.339 \\
 & Two-infall & 0.119 & 0.481 & 0.812 & 1.281 \\
\hline
\end{tabular}
\unskip\label{output/scores_table.tex}\unskip%

\end{table*}

Table \ref{tab:scores} presents the quantitative scores which measure the difference between the multi-zone outputs and APOGEE data for four observables. Details of the calculations for each observable are presented in the corresponding subsections of Section \ref{sec:multizone-results}, but in summary, divergence statistics between the multi-zone output and APOGEE data (Equation \ref{eq:kl-divergence} for the [Fe/H] and [O/Fe] DFs, Equation \ref{eq:2d-kl-divergence} for the [O/Fe]--[Fe/H] plane, and Equation \ref{eq:age-ofe-score} for the age--[O/Fe] plane) are computed within each Galactic region. The overall score for the model is the average of the divergence statistics across all regions, weighted by the number of APOGEE targets in each region. The scores can be strongly affected by zero-point offsets between the model and observed abundance distributions, so we caution against using these scores as an absolute metric of model performance.

\bibliography{bib}

\end{document}